%% file: CPTrack.tex
\documentclass[12pt]{report}
\usepackage{amsmath,amssymb,amsfonts,amscd,theorem,graphics}
\usepackage[latin1]{inputenc}
\input{cmds}

\def\bbf{\bfseries\boldmath}

\def\Bi{\begin{itemize}}
\def\Ei{\end{itemize}}

\def\sum{\mathop{\Sigma}\limits}
\def\prod{\mathop{\Pi}\limits}

\def\bigcup{\mathop{\cup}\limits}

\def\bplus{\mathop{\boxplus}\limits}

\def\To{\longrightarrow }

\def\CC{{\mathbb{C}\,}}
\def\NN{{\mathbb{N}\,}}

\def\Repsp{\operatorname{Repsp}}

\def\TAN{\operatorname{TAN}}

\def\RxL{_{R\times L}}

\def\GL{\operatorname{GL}}

\def\bM{\mbox{\boldmath ${M}$\,}}
\def\tt{{}^{\text{\dag}}}

{\theorembodyfont{\rmfamily}
  }
{\theorembodyfont{\rmfamily}
  }

{\theorembodyfont{\rmfamily}
  }
{\theorembodyfont{\rmfamily}
  }

{\theorembodyfont{\rmfamily}
  }

\def\lr{left (resp. right) }

\newcommand{\gfrak}{\mathfrak{g}}

\begin{document}

\include{CPTrack00}

\pagestyle{myheadings}

\def\thepage{\arabic{page}}

\include{CPTrack1}
\include{CPTrack2}

\include{CPTrack3}

\include{CPTrack4}

\include{auteurs}


\end{document}

%% file: cmds.tex
\def\simarrow{\mathrel{\raise -0.5mm\hbox{$\sim$}}\hspace{-1.8mm}{\rightarrow} } 

\def\bsimarrow{\leftarrow\hspace{-0.7mm}\mathrel{\raise -0.5mm\hbox{$\backsim$}} }

\def\bt{\begin{tabular}}
\def\te{\end{tabular}}

\def\lettrine#1#2#3{\noindent\hangindent#1\hangafter-#2
\hskip-#1\smash{\hbox to #1{#3\hfill}}\ignorespaces}

\newcommand{\To}[1]{\mathop{\to}\limits_{#1}}

\def\BM{\begin{pmatrix}}
\def\EM{\end{pmatrix}}

\def\txt{\textstyle}

\def\d=f{\buildrel\hbox{\scriptsize d\'{e}f}\over \Longleftrightarrow}

\def\square{\hfill\hbox{\vrule height .9ex width .8ex depth -.1ex}}
\def\rit{\text{\it I\hskip -2pt  R}}

\def\Bd{{\text B}}

\def\Ed{{\text E}}

\def\Ms{{\cal M}}

\def\Ps{{\cal P}}

\def\be{\begin{equation}}
\def\ee{\end{equation}}
\def\beqn{\begin{eqnarray}}
\def\eeqn{\end{eqnarray}}
\def\nobeqn{\begin{eqnarray*}}
\def\noeeqn{\end{eqnarray*}}
\def\ba{\left(\begin{array}}
\def\ea{\end{array} \right) }

\def\bpr{\paragraph{Proof.}}

\def\epr{\square\vskip 6pt}

\def\o{\overline}
\def\and{\; \mbox{and} \;}

\newcommand{\half}{\frac{1}{2}}

\def\hfl#1#2{\smash{\mathop{\hbox to 12mm{\rightarrowfill}}
\limits^{\scriptstyle #1}_{\scriptstyle #2}}}

\def\Be{\begin{enumerate}}
\def\Ee{\end{enumerate}}

\def\Bena{\begin{enumerate}
\def\labelenumi{\theenumi)}
\def\theenumi{\arabic{enumi}}
\def\labelenumii{\theenumii)}
\def\theenumii{\alph{enumii}}}

\def\Bean{\begin{enumerate}
\def\labelenumii{\theenumii)}
\def\theenumii{\arabic{enumii}}
\def\labelenumi{\theenumi)}
\def\theenumi{\alph{enumi}}}

\def\Bero{\begin{enumerate}
\def\labelenumii{\theenumii)}
\def\theenumii{\arabic{enumii}}
\def\labelenumi{(\theenumi)}
\def\theenumi{\roman{enumi}}}

\def\BeRo{\begin{enumerate}
\def\labelenumii{\theenumii)}
\def\theenumii{\arabic{enumii}}
\def\labelenumi{(\theenumi)}
\def\theenumi{\Roman{enumi}}}

\def\Bi{\vskip 11pt\begin{itemize}\itemsep=18pt}
\def\Ei{\end{itemize}\vskip 11pt}

\def\Bd{\begin{description}}
\def\Ed{\end{description}}

\def\R{\right}
\def\L{\left}
\def\F{\frac}

%% file: CPTrack00.tex
\thispagestyle{empty}

\null\vfill
\begin{center}
{\LARGE  A new track for unifying general relativity with quantum field theories}%
 \vfill
{\sc C. Pierre\/}
\vskip 11pt

Institut de Mathématique pure et appliquée\\
Université de Louvain\\
Chemin du Cyclotron, 2\\
B-1348 Louvain-la-Neuve,  Belgium\\
pierre@math.ucl.ac.be

\vfill

{\bfseries Abstract}
\vskip 11pt
\end{center}

\begin{quote}

In the perspective of unifying quantum field theories with general relativity, the equations of the internal dynamics of the vacuum and mass structures of a set of interacting particles are proved to be in one-to-one correspondence with the equations of general relativity.

This leads to envisage a high value for the cosmological constant, as expected theoretically.
 \end{quote}\vfill\eject


%% file: CPTrack1.tex
\chapter{Introduction}

\thispagestyle{empty}

\addtocontents{toc}{\protect\thispagestyle{empty}}

It is generally believed that a convincing theory of quantum gravity \cite{Ash}, \cite{Alv}, \cite{Dew}, \cite{Wal}, \cite{Whe} will emerge from some unification of quantum field theory with general relativity.

This objective motivated the creation of the standard perturbative approaches based on Feynman perturbation theory for graviton modes of the form $g_{\mu \nu }=\eta _{\mu \nu }+h_{\mu \nu}$ where $h_{\mu \nu }$ refers to a small excitation of a flat metric \cite{Pag}, \cite{E-G-H}, \cite{Hoo2}.  But, these models failed because they are perturbatively nonrenormalizable \cite{Nie}, \cite{Gro}, \cite{D-V}.

In order to overcome these problems, string theories \cite{G-S-W}, \cite{Pol}, \cite{Wit2} and loop quantum gravity \cite{R-S}, \cite{Smo}, \cite{Bae} were created in such a way that point like models of elementary particles be replaced by one-dimensional models.

In spite of an intense activity in this field and evident successes, the contact with experiment still remain nebulous and, furthermore, the conceptual framework of (super)string theories has not reached the expected maturity degree.

So, as G. 't Hooft \cite{Hoo1} pointed out, it might be that quantum gravity could not be solved without revising the principles of quantum theory and, especially, at the Planck scale.

As a matter of fact, it is commonly admitted that  Lorentz invariance could be broken at the Planck length due to the strong fluctuations of the space-time at this scale \cite{H-S-L-S}, \cite{Wit1}.

To go beyond these difficulties, it was envisaged in \cite{Pie3} and in \cite{Pie5} {\bf to enlarge the conceptual framework of quantum field theories\/} in order that the new proposed quantum model takes into account the quantum Physics below the Planck scale and corresponds to quantum field theories at the scale for which they were developed \cite{Ati}.

But, to become acceptable, the enlargement of the conceptual frame of quantum field theories has to imply some unification of these with the equations of general relativity.

The problem is that general relativity is a classical theory having resisted until now to a convincing quantization.

{\bbf It is thus the aim of this paper to reconsider the matter in question on new basis at the light of what can bring close together general relativity and quantum field theories.

What characterizes general relativity with respect to a possible quantization is:}
\Be
\item The influence of the space-time curvature on the matter and vice versa.

\item The existence of the cosmological constant $\lambda $ interpreted by Y. Zel'dovich \cite{Zel1} as a medium tied to the vacuum polarization of the QFT (quantum field theory).
\Ee

Reciprocally, one of the main features of QFT which could be used to quantize GR (General relativity) consists in the creation of (pairs of) particles from the vacuum energy which reacts backwards on the curvature of the space-time: this means that the energy of the created particles is taken away from the space-time itself \cite{Dew}.

In this respect, as the quantum level is essentially that of elementary particles, it was envisaged \cite{Pie1} by the author to set up equations describing the (internal) dynamics of a set of elementary particles in such a way that these equations be isomorphic to the equations of general relativity with respect to a metric contraction given by a suitable compactification from the quantum level to the classical level and associated with the condition $\hbar\to 0$~.

To fulfil these conditions {\bbf in order to bridge the gap between QFT and GR\/}, it was envisaged that {\bbf the expanding space-time\/}, to which the cosmological constant of GR may correspond, {\bbf could constitute the fundamental structure of the QFT vacuum shared out amongst the considered elementary particles.}

In this context, every elementary particle would then be characterized by an internal expanding space-time structure constituting its own vacuum from which its matter shell could be generated due to the strong fluctuations of the internal space-time of its internal vacuum.

Indeed, at this scale which likely corresponds to the Planck scale, the fluctuations of the space-time generate local strong curvatures which are responsible for the generation of degenerate singularities allowing by versal deformations and blowups of these to create two external contracting enveloping structures above the particle internal vacuum in such a way that the most external structure is interpreted as its mass shell.

So, an important part of the vacuum of QFT would be of space-time nature distributed amongst {\bbf the internal vacua of the elementary particles\/} constituting massless particles {\bbf potentially able to generate their mass shells\/} due to the strong fluctuations of their internal space-time vacua.

Thus, the attempt of unifying QFT with GR leads us to {\bbf envisage elementary particles endowed with internal structures which must be of space-time type and quantized.}

The first step, developed in chapter 1, then consists in {\bbf generating mathematically space and time internal fields that must be really quantized\/}: this can be only realized algebraically by assuming that {\bbf quanta must have an algebraic structure given by algebraic closed subsets\/} characterized by a Galois extension degree $N$~, $N\in\NN$~.

As extensively developed in \cite{Pie5}, the internal fields of elementary particles have {\bbf a twofold nature\/} due to the fact that {\bbf every elementary particle has to be viewed as an elementary bisemiparticle\/} composed of a left semiparticle, localized in the upper half space, and of a right symmetric (co)semiparticle, localized in the lower half space in such a way that:
\Bean
\item the right semiparticle, dual to the left semiparticle, is observed as projected on the latter and is thus normally unobservable.
\item the product of the internal structure of the right semiparticle by its left equivalent gives rise to a ``working interaction space'' responsible for the electric charge and the magnetic moment.
\Ee

In this respect, the internal space-time structure of the vacuum of a bisemiparticle will be composed of an internal time field and of an internal space field having a bilinear nature and being localized in orthogonal spaces.

{\bbf The time internal field (as well as the space internal field) is then given mathematically by a (bisemi)sheaf of differentiable bifunctions over an algebraic bilinear semigroup\/}: a bifunction is defined as the product of a right function localized in the lower half space by its symmetric left equivalent and the considered algebraic bilinear semigroup is $\GL_2(F_{\o v}\times F_v)$ covering its linear correspondent \cite{Pie2} and defined over the product $(F_{\o v}\times F_v)$ of the set of pairs $\{F_{\o v_{\mu ,m_\mu }},F_{v_{\mu ,m_\mu }}\}$ or right and left real ramified algebraic subsets having a structure at $\mu $ quanta of degree
\[ [F_{\o v_{\mu ,m_\mu }}:K] = [F_{v_{\mu ,m_\mu }}:K]=\mu \centerdot N\]
over a global number field $K$ of characteristic 0~.

The representation space $\Repsp(\GL_2(F_{\o v}\times F_v))$ of $\GL_2(F_{\o v}\times F_v)$ then consists in the product, right by left, of {\bbf two symmetric towers of conjugacy class representatives\/} in such a way that the functional representation subspaces of the products of {\bbf the corresponding right and left compactified conjugacy class representatives behave like harmonic oscillators.}  This leads us to consider that {\bbf the functional (modular) representation space of $\GL_2(L_{\o v}\times L_v)$\/}, where $L_{\o v}$ (resp. $L_v$~) results from the compactification of $F_{\o v}$ (resp. $F_v$~), {\bbf can be  interpreted as a time internal field of an elementary bisemiparticle.}

The corresponding internal space field can be obtained from the internal time field throughout a smooth (bi)endomorphism based on Galois antiautomorphisms as developed in proposition 2.7.  The end of chapter 1 deals with {\bbf the compactification of the conjugacy class representatives of the ``time'' and ``space'' bilinear algebraic semigroups\/} leading to continuous upper and lower 4-dimensional (semi)manifolds of space-time.

So, {\bbf the consideration of algebraic quanta at a sub-Planckian scale allows to generate time and space quantized internal fields of the internal vacua\/} of elementary particles likely at the Planck scale.

These time and space internal fields are products, right by left, of time and space semifields restricted to the lower and upper half spaces and described mathematically by (semi)sheaves of differentiable functions over the appropriate algebraic bilinear semigroups.

Due to the strong fluctuations of space-time at this level, these differentiable functions become afflicted by degenerate singularities of which versal deformations and blowups allow {\bbf to generate two covering ``middle-ground'' and ``mass'' fields of space and time.}

The internal vacuum of an elementary (bisemi)particle is thus described mathematically by a space and time internal field and by an enveloping ``middle-ground'' space and time field.  A process of versal deformations and blowups of the singularities on the ``middle-ground'' space and time fields is responsible for the creation or {\bbf generation of the enveloping ``mass'' fields of space and time, phenomenon corresponding to the creation of this (bisemi)particle from its own vacuum.}

The generation of the middle-ground and mass fields of an elementary particle from its internal space and time fields constitute the contents of chapter 3, which allows to formulate {\bbf the equations of the internal dynamics of a bisemifermion.}

On the other hand, the unification of quantum field theory with general relativity, leading to a coherent quantum gravity theory, requires {\bf a new interpretation of the equations of general relativity at the Planck scale\/}.

These objectives will be reached if it is taken into account that:

\Be
\item an {\bf  initial state  mathematically well-defined\/} must be introduced ``at the beginning'' in such a way that:
\Be
\item it corresponds to some {\bf universal structure\/} in the perspective of the Langlands program by a one-to-one correspondence with its automorphic representation.
\item it be of (bilinear) space-time nature.
\item its {\bf space-time structure be totally quantized\/} from an algebraic point of view.
\item it gives rise to the fundamental structure of the {\bf internal vacua of the elementary particles\/}.
\item it be directly {\bbf connected to the (small) value of the cosmological constant $\Lambda $ \/} \cite{Wei2}.
\Ee

\item {\bf the interactions between elementary particles\/} must  appear naturally (i.e. without divergences) in the model by considering:
\Be
\item bilinear interactions between pairs of semiparticles as it was developed in chapter 5 of \cite{Pie5}.
\item a unification of electro-magnetism with gravitation as it was hoped by Einstein \cite{Ein2}.
\Ee
\Ee

In this new context, a special importance is thus given to the structure of elementary particles whose space-time nature is quantized and less to the evolution of states of these particles as currently done in quantum field theories and (super)string theories.
\vskip 11pt

The idea then consists in building up a quantum gravity theory at the level of elementary particles in such a way that:

\Be
\item {\bf the space-time structure\/}, constituting the ``initial'' state {\bf of each elementary particle\/}, corresponds to its internal vacuum structure of expanding nature from which its mass structure of contracting nature can be generated dynamically from the singularities on its internal vacuum structure submitted to strong fluctuations at the Planck scale.

\item {\bf the set of equations of vacuum and mass structure\/} of elementary particles {\bf is in one-to-one correspondence with the equations of general relativity\/} if:
\Bi
\item the big points, describing the quanta, at the Planck scale are viewed as ordinary points at the classical level by means of a metric contraction based on a condition equivalent to $\hbar \to 0$~.
\item the one-dimensional components of the time and space structures of the elementary particles, considered as flow lines, are compactified to give rise to a four dimensional manifold endowed with a Riemann geometry.
\Ei
\Ee

The equations of general relativity, generated in that manner at the Planck scale, are characterized by a {\bf high value of the cosmological constant\/}, corresponding to the set of internal vacua of elementary particles, in contrast with the standard interpretation of these equations where the cosmological constant, being very small \cite{Edd}, is used to adjust the geometry in order that it be on a average flat.




Chapter 4 is then devoted to the set up of the equations of the internal dynamics of a set of interacting particles and to the proof of their equivalence with the equations of general relativity in the perspective of unifying quantum field theories with general relativity.

All developments of this paper refer to the preprint ``Algebraic quantum theory'' \cite{Pie5}.

%% file: CPTrack2.tex
\chapter[The space-time structure of the internal vacua of elementary (bisemi)- fermions]{The space-time structure of the internal vacua of elementary (bisemi)fermions}

\thispagestyle{empty}

A first step will then consist in recalling the algebraic quantized space-time structure of the vacuum of elementary particles, under the circumstances the elementary (stable) fermions, i.e.
\Bi
\item the leptons $e^-$~, $\mu ^-$~, $\tau ^-$ and their neutrinos.
\item the quarks $u ^+$~, $d^-$~, $s^-$~, $c^+$~, $b^-$~, $t^+$~.
\Ei

\section{An elementary bisemifermion}

As developed in \cite{Pie1} and \cite{Pie3}, {\bf an elementary fermion must be viewed as an elementary bisemifermion}:
\Bi
\item composed of (the product of) a left semifermion, localized in the upper half space, and of a right semifermion, localized in the symmetric lower half space.

\item centred on an emergence point allowing the transfer of quanta from the time structures of a semifermion to their space structures and vice-versa.

\item to which it can be associated a ``working space'' composed of (tensor) products between right and left (time and space) internal structures, respectively of a right and of a left semifermion.
\Ei

\section{``Initial state'' of a bisemifermion}
The ``initial state'' of a bisemifermion, corresponding to its most internal vacuum structure, must be chosen to be:
\Bean 
\item of mathematical nature, and, consequently, of space-time type.
\item discrete, and, thus quantized.
\Ee

The only way to satisfy these two conditions is that the internal vacuum structure of a bisemifermion be {\bf of algebraic type\/}.

And, as the bilinear nature of a bisemifermion must be taken into account, an algebraic bilinear semigroup over a set of products, right by left, of symmetric algebraic subsets characterized by increasing Galois extension degrees was introduced in \cite{Pie1} in such a way that a natural automorphic representation corresponds to it by means of a Langlands global correspondence \cite{Pie2}.

\section{Algebraic bilinear semigroup over symmetric splitting fields}

\Bi
\item Let $K$ be a global number field of characteristic 0 and let $K[x]$ denote a polynomial ring composed of a set of pairs of polynomials $\{P(x),P(-x)\}$~, $x$ being a time or space variable.

The real and complex algebraic extensions of $K$~, noted respectively  $F_{\o v-v}$ and $F_{\o \omega -\omega }$~, are assumed to be {\bf symmetric splitting fields\/} $F_{\o v-v}=F_{\o v}\cup F_v$ and 
$F_{\o \omega -\omega }=F_{\o \omega }\cup F_\omega $ composed of right extension semifields $F_{\o v}$ and $F_{\o\omega }$ and of left extensions semifields $F_v$ and $F_\omega $ in one-to-one correspondence \cite{Pie1}.
\vskip 11pt

\item {\bf The real ramified algebraic subsets\/} are assumed to be generated from {\bf irreducible\/} (one-dimensional) {\bf algebraic closed subsets\/} characterized by a Galois extension degree $[F_{v^1_\mu }:K]\equiv [F_{\o v^1_\mu }:K]=N$~, $1\le\mu \le q\le\infty $~, equal to $N$ and {\bf interpreted as space or time quanta\/}.

The real ramified algebraic subsets $F_{v_\mu }$ (resp. $F_{\o v_\mu }$~) are indexed in equivalence classes characterized by their ranks (i.e., their Galois extension degrees) which are integers modulo $N$~:
\[ [F_{v_\mu }:K]=*+\mu \centerdot N \qquad \text{(resp.} \quad
[F_{\o v_\mu }:K]=*+\mu \centerdot N\ )\]
where:
\Bi
\item $F_{v_\mu }\subset F_v$~, $1\le \mu \le q\le\infty $ \quad (resp. 
$F_{\o v_\mu }\subset F_{\o v}$~, $1\le \mu \le q\le\infty $~).
\item $*$ denotes an integer inferior to $N$~.
\Ei

The ranks will be generally chosen to be equal to 0 modulo $N$~.  So, we get {\bf a tower of classes of real ramified algebraic subsets\/}:
\begin{align*}
 F_{v_1}&\subset \cdots \subset F_{v_{\mu ,m_\mu }} \subset \cdots \subset F_{v_{q,m_q}} \\
 \text{(resp.} \quad 
F_{\o v_1}&\subset \cdots \subset F_{\o v_{\mu ,m_\mu }} \subset \cdots \subset F_{\o v_{q,m_q}} \ ),\end{align*}
where $ F_{v_{\mu ,m_\mu }} $ (resp. $ F_{\o v_{\mu ,m_\mu }} $~) is the $m_\mu $-th representative of the $\mu $-th equivalence class, \\
in such a way that {\bbf all representatives of the $\mu $-th equivalence class have a structure at $\mu $ quanta\/} of degree $N$~, since their extension degrees (or ranks) are
\[ [F_{v_{\mu ,m_\mu }}:K]=\mu \centerdot N \qquad \text{(resp.} \quad
[F_{\o v_{\mu ,m_\mu }}:K]=\mu \centerdot N \ ).\]
It must be noted that the real ramified algebraic subsets are included, in the sense of \cite{Pie3}, into the corresponding complex ramified algebraic subsets.
\vskip 11pt

\item The next step consists in considering an algebraic group over the product, right by left, of corresponding equivalent representatives of
\begin{align*}
 F_{v}&=\{F_{v_1},\cdots,F_{v_{\mu ,m_\mu }},\cdots,F_{v_{q,m_q}}\}\\
\text{and of} \quad 
 F_{\o v}&=\{F_{\o v_1},\cdots,F_{\o v_{\mu ,m_\mu }},\cdots,F_{\o v_{q,m_q}}\}
 \end{align*}
 to take into account the bilinear (or twofold) structure of the elementary bisemiparticles as noticed in section 2.2.
 
 Let then $T_2(F_v)$ (resp. $T^t_2(F_{\o v})$~) denote the (semi)group of upper (resp. lower) triangular matrices of order 2 over $F_v$ (resp. $F_{\o v}$~).
 
 An {\bf algebraic bilinear general semigroup\/} 
 \[ \GL_2(F_{\o v}\times F_v) \equiv T^t_2(F_{\o v}) \times T_2(F_v)\]
 can be introduced so that:
 \Bean
 \item the product $(F_{\o v}\times F_v)$ is taken over the set of corresponding pairs $
 \{F_{\o v_{\mu ,m_\mu }},\linebreak F_{v_{\mu ,m_\mu }}\}_{v_{\mu ,m_\mu }}$ of right and left real ramified algebraic subsets.
 
 \item $\GL_2(F_{\o v}\times F_v)$ has the Gauss bilinear decomposition:
 \[ \GL_2(F_{\o v}\times F_v) = [D_2(F_{\o v})\times D_2(F_v)][UT^t_2(F_{\o v})\times UT_2(F_v)]\]
 where:
 \Bi
 \item $D_2(\centerdot)$ is the subgroup of diagonal matrices.
 \item $UT_2(\centerdot)$ (resp. $UT^t_2(\centerdot)$~) is the subgroup of upper (resp. lower) unitriangular matrices.
 \Ei
 
 \item $\GL_2(F_{\o v}\times F_v)$ covers its linear equivalent $\GL_2(F_{\o v}- F_v)$~.
 
 \item the modular representation space $\Repsp (\GL_2(F_{\o v}\times F_v))$ of
 $\GL_2(F_{\o v}\times F_v)$ is given by the tensor product $M_R(F_{\o v})\otimes M_L(F_v)$ of a right $T^t_2(F_{\o v})$-semimodule $M_R(F_{\o v})$ by a left $T_2(F_v)$-semimodule $M_L(F_v)$~.

\item the $\mu $-th conjugacy class representative
$\GL_2(F_{\o v_{\mu ,m_\mu }}\times F_{v_{\mu ,m_\mu }})$ of
$\GL_2(F_{\o v}\times F_v)$~, with respect to the product, right by left,
$(F_{\o v^1_\mu }\times F_{v^1_\mu })$ of irreducible algebraic closed subsets of rank $N$~ , has for representation the $\GL_2(F_{\o v_{\mu ,m_\mu }}\times F_{v_{\mu ,m_\mu }})$-subbisemimodule $(M_{F_{\o v_{\mu ,m_\mu }}}\otimes M_{F_{v_{\mu ,m_\mu }}})$~.
\Ee
\Ei

\section[The compactification of the algebraic bilinear semigroup $\GL_2(F_{\o v }\times F_{v})$]{\bbf The compactification of the algebraic bilinear semigroup $\GL_2(F_{\o v }\times F_{v})$}

\Bi
\item The problem is that the conjugacy class representatives of $\GL_2(F_{\o v }\times F_{v})$
are not locally compact.

Now, the two elements $M_{F_{\o v_{\mu ,m_\mu }}}$ and $M_{F_{v_{\mu ,m_\mu }}}$ of each conjugacy class representative 
$\GL_2(F_{\o v_{\mu ,m_\mu }}\times F_{v_{\mu ,m_\mu }})$ of $\GL_2(F_{\o v }\times F_{v})$ rotate in opposite senses giving them a spin orientation according to \cite{Pie5}.

The rotation of the elements $M_{F_{\o v_{\mu ,m_\mu }}}$ and $M_{F_{v_{\mu ,m_\mu }}}$ of the conjugacy class representatives leads to a compactification of them as developed by W. Fulton and R. McPherson \cite{F-M} and adapted in the present case in \cite{Pie2}.
\vskip 11pt

\item {\bbf The compactification of 
$M_{F_{v_{\mu ,m_\mu }}}$ 
(resp. $M_{F_{\o v_{\mu ,m_\mu }}}$~)\/} given by the map:
\[
\gamma ^c_{M_{F_{v_{\mu ,m_\mu }}}}: \quad M_{F_{v_{\mu ,m_\mu }}} \To M_{L_{v_{\mu ,m_\mu }}}\qquad
\text{(resp.} \quad 
\gamma ^c_{M_{F_{\o v_{\mu ,m_\mu }}}}: \quad M_{F_{\o v_{\mu ,m_\mu }}} \To M_{L_{v_{\mu ,m_\mu }}}\ )
\]
results from:
\Bean
\item a compactification of the $\mu $ irreducible algebraic closed subsets, i.e. of the $\mu $ quanta, of $M_{F_{v_{\mu ,m_\mu }}}$ (resp. $M_{F_{\o v_{\mu ,m_\mu }}}$~) by a sequence of blowups on the algebraic points of each irreducible subset transforming it into an irreducible completion centred on a point in the upper (resp. lower) half space according to the maps:
\[ \gamma ^c_{v^1_\mu }: F_{v^1_\mu }\To L_{v^1_\mu }
\qquad \text{(resp.} \quad 
\gamma ^c_{\o v^1_\mu }: F_{\o v^1_\mu }\To L_{\o v^1_\mu }\ )\]
where:
\Bi
\item the closed irreducible algebraic subset
$F_{v^1_\mu }$ (resp. $F_{\o v^1_\mu }$~) is one of the $\mu $ quanta of 
$M_{F_{\o v_{\mu ,m_\mu }}}$ (resp. $M_{F_{v_{\mu ,m_\mu }}}$~).
\item $L_{v^1_\mu }$ (resp. $L_{\o v^1_\mu }$~) is an irreducible completion of rank $N$~, corresponding to a compactified quantum.
\Ei

\item The connectedness of the set of the $\mu $ irreducible completions put from end to end alter their compactifications in such a way that they generate a one-dimensional closed string
$M_{L_{v_{\mu ,m_\mu }}}$ (resp. $M_{L_{\o v_{\mu ,m_\mu }}}$~).
\Ee
\vskip 11pt

\item By this way, each compactified quantum $L_{v^1_\mu }$ (resp. $L_{\o v^1_\mu }$~) on the conjugacy class representative $M_{L_{v_{\mu ,m_\mu }}}$ (resp. $M_{L_{\o v_{\mu ,m_\mu }}}$~), which is a closed string, can be viewed as one of its $\mu $ {\bf quantum ``big points''\/} centred on the compactification points of the blowups as described in (a).
\Ei

\section[The compactified algebraic bilinear semigroup $\GL_2(L_{\o v}\times L_v)$]{\bbf The compactified algebraic bilinear semigroup\protect\newline $\GL_2(L_{\o v}\times L_v)$}

\Bi
\item So, the compactification of all the conjugacy class representatives
$(M_{F_{\o v_{\mu ,m_\mu }}} \otimes M_{F_{v_{\mu ,m_\mu }}})$ of the bilinear algebraic semigroup $\GL_2(L_{\o v}\times L_v)$ transforms it into the bilinear algebraic semigroup
$\GL_2(L_{\o v}\times L_v)$ where
\begin{align*}
L_v &= \{L_{v_1},\cdots, L_{v_{\mu ,m_\mu }},\cdots,L_{v_{q,m_q}}\}\\
\text{(resp.}\quad 
L_{\o v} &= \{L_{\o v_1},\cdots, L_{\o v_{\mu ,m_\mu }},\cdots,L_{\o v_{q,m_q}}\}\ )\end{align*}
is the set of completions corresponding to the set $F_v$ (resp.
 $F_{\o v}$~) of real ramified algebraic subsets.
 \vskip 11pt
 
\item The $\mu $-th conjugacy class representative $\GL_2(L_{\o v_{\mu ,m_\mu }} \times L_{v_{\mu ,m_\mu }})$ of $\GL_2(L_{\o v}\times L_v)$~, with respect to the product, right by left, $(L_{\o v^1_\mu }\times L_{v^1_\mu })$ of irreducible completions of rank $N$~, has for representation the $\GL_2(L_{\o v_{\mu ,m_\mu }} \times L_{v_{\mu ,m_\mu }})$-subbisemimodule
$(M_{L_{\o v_{\mu ,m_\mu }}}\otimes M_{L_{v_{\mu ,m_\mu }}})$~.
\vskip 11pt

\item But, as the elements $M_{L_{\o v_{\mu ,m_\mu }}}$ and $M_{L_{v_{\mu ,m_\mu }}}$ of
$\GL_2(L_{\o v}\times L_v)$ rotate in opposite senses as noticed in section 2.4, it is more exactly the Lie algebra $\gfrak\ell_2(L_{\o v}\times L_v)$ of the bilinear algebraic semigroup
$\GL_2(L_{\o v}\times L_v)$ which would be considered.
\Ei

\section{Proposition}

{\em The discrete structure, generated by the algebraic bilinear semigroup
$\GL_2(F_{\o v}\times F_v)$~:
\Bi
\item corresponds to the time structure of the internal vacuum of an elementary (bisemi)\-fermion.
\item has been transformed by the compactification map:
\[ \gamma ^c_{F_{\o v}\times F_v}: \quad
\Repsp(\GL_2(F_{\o v}\times F_v)) \To \Repsp (\GL_2(L_{\o v}\times L_v))\]
into a corresponding {\bf locally compact quantized time structure\/}.\Ei}
\vskip 11pt

\bpr
\Be\item It was proved in \cite{Pie5} and in \cite{Pie3} that $\Repsp (\GL_2(F_{\o v}\times F_v))$ corresponds to the time structure of the internal vacuum of an elementary bisemifermion because its conjugacy class representatives $(M_{F_{\o v_{\mu ,m_\mu }}}\otimes
M_{F_{v_{\mu ,m_\mu }}})$ are isomorphic, under the compactifications $\gamma ^c_{M_{F_{\o v_\mu }}}\times \gamma ^c_{M_{F_{v_\mu }}}$~, to products of pairs of strings
$(M_{L_{\o v_{\mu ,m_\mu }}}\otimes
M_{L_{v_{\mu ,m_\mu }}})$ behaving like harmonic oscillators.  And, a (bisemi)sheaf of differentiable (bi)functions on it is a (time) field as it will be seen.

\item As $M_{F_{v_{\mu ,m_\mu }}}$ (resp. $
M_{F_{\o v_{\mu ,m_\mu }}}$~) is composed of $\mu $ quanta, its compactified equivalent
$M_{L_{v_{\mu ,m_\mu }}}$ (resp. $
M_{L_{\o v_{\mu ,m_\mu }}}$~) will also have a structure at $\mu$ quanta set out in a continuous way.\epr
\Ee

\section{Proposition}

{\em Under the composition of maps $\gamma _{t\RxL\to r\RxL}\circ E\RxL$~, where:
\Bi
\item $E\RxL: \Repsp(\GL_2(L_{\o v}\times L_v)_t) \to \Repsp (\GL^*_2(L^*_{\o v}\times L^*_v)_t) 
\oplus \Repsp (\GL^I_2(L^I_{\o v}\times L^I_v)_t)$ is a smooth biendomorphism
\item $\gamma _{t\RxL\to r\RxL}$ sends ``time'' complementary bistructures into orthogonal ``space'' bistructures,
\Ei
the ``time'' representation space $\Repsp (\GL_2(L_{\o v}\times L_v)_t)$ can be transformed into:
\Bean
\item a reduced ``time'' representation space $\Repsp (\GL^*_2(L^*_{\o v}\times L^*_v)_t)$ over a set of reduced completions $(L^*_{\o v}\times L^*_v)$~.
\item a complementary ``space'' representation space $\Repsp (\GL^I_2(L^I_{\o v}\times L^I_v)_r)$ over a complementary set of completions $(L^I_{\o v}\times L^I_v)$~, localized in a space perpendicular to\linebreak $\Repsp (\GL^*_2(L^*_{\o v}\times L^*_v)_t)$~.
\Ee}
\vskip 11pt

\bpr
\Be
\item It was seen in \cite{Pie5} that the ``time'' representation space
$\Repsp (\GL_2(L_{\o v}\times L_v)_t)$~, submitted to Galois antiautomorphisms on its right and left parts, can be decomposed into the direct sum of two non connected ``time'' representation spaces of bilinear algebraic semigroups:
\[ E\RxL: \quad \Repsp (\GL_2(L_{\o v}\times L_v)_t)\To
\Repsp (\GL^*_2(L^*_{\o v}\times L^*_v)_t) \oplus
\Repsp (\GL^I_2(L^I_{\o v}\times L^I_v)_t)\]
corresponding to a smooth biendomorphsim $E\RxL$ in such a way that the biquanta (i.e. products of pairs of corresponding quanta on the same conjugacy class representatives
$(M_{L_{\o v_{\mu ,m_\mu }}}\otimes M_{L_{v_{\mu ,m_\mu }}})$~), extracted from
$\Repsp (\GL^*_2(L^*_{\o v}\times L^*_v)_t)$ by Galois antiautomorphisms, be used to build up a new non connected complementary ``time'' representation space
$\Repsp (\GL^I_2(L^I_{\o v}\times L^I_v)_t)$~.

\item Under the map
\[ \gamma _{t\RxL\to r\RxL}: \quad
\Repsp (\GL^I_2(L^I_{\o v}\times L^I_v)_t)\To
\Repsp (\GL^I_2(L^I_{\o v}\times L^I_v)_r)\]
the complementary `` time'' representation space
$\Repsp (\GL^I_2(L^I_{\o v}\times L^I_v)_r)$ is sent, out of the origin, biquantum by biquantum, into the orthogonal space generating by this way the complementary representation space $\Repsp (\GL^I_2(L^I_{\o v}\times L^I_v)_r)$ of ``space''.

\item Thus, the composition of maps:
\begin{multline*}\gamma _{t\RxL\to r\RxL}\circ E\RxL: \quad
\Repsp (\GL_2(L_{\o v}\times L_v)_t)\\ \To
\Repsp (\GL^*_2(L^*_{\o v}\times L^*_v)_t)\oplus
\Repsp (\GL^I_2(L^I_{\o v}\times L^I_v)_r)\end{multline*}
transforms the ``time'' representation space
$\Repsp (\GL_2(L_{\o v}\times L_v)_t)$ into the reduced ``time'' representation space
$\Repsp (\GL^*_2(L^*_{\o v}\times L^*_v)_t)$ and into the complementary ``space'' representation space $\Repsp (\GL^I_2(L^I_{\o v}\times L^I_v)_r)$~.
\Ee

As a consequence, {\bf the ``time'' structure of the internal vacuum\/} of an elementary (bisemi)fermion, given by the ``time'' representation space
$\Repsp (\GL_2(L_{\o v}\times L_v)_t)$~, {\bf can be transformed\/} partially, or completely, {\bf into a ``space'' structure\/}, given by the complementary ``space'' representation space
$\Repsp (\GL^I_2(L^I_{\o v}\times L^I_v)_r)$ {\bf and this ``space-time'' structure is quantized\/}.\epr

\section[Bisemisheaf of differentiable (bi)functions on $\GL_2(L_{\o v}\times L_v)$]{\bbf Bisemisheaf of differentiable (bi)functions on\protect\newline $\GL_2(L_{\o v}\times L_v)$}

\Bi
\item Let $\phi _L(M_{v_\mu })$ (resp. $\phi _R(M_{\o v_\mu })$~) denote a complex-valued differentiable function over the $\mu $-th real conjugacy class representative $M_{L_{v_\mu }}$ (resp. $M_{L_{\o v_\mu }}$~) of $T_2(L_v)$ (resp. $T^t_2(L_{\o v})$~) $\subset \GL_2(L_{\o v}\times L_v)$ and let $\phi _R(M_{\o v_\mu })\otimes \phi _L(M_{v_\mu })$ be the corresponding bifunction on the conjugacy class representative $(M_{L_{\o v_\mu }}\otimes M_{L_{v_\mu }})$ of $\GL_2(L_{\o v}\times L_v)$~.
\vskip 11pt

\item The set $\{\phi _L(M_{v_{\mu ,m_\mu }})\}_{\mu, m_\mu }$ (resp.
$\{\phi _R(M_{\o v_{\mu ,m_\mu }})\}_{\mu, m_\mu }$~) of $\CC-$valued differentiable functions, localized in the upper (resp. lower) half space and defined over the $T_2(L_v)$ (resp. $T^t_2(L_{\o v})$~)-semimodule $M_L(L_v)$ (resp. $M_R(L_{\o v})$~), constitutes the set
$\Gamma (\phi _L(M_L(v)))$ (resp. $\Gamma (\phi _R(M_R(L_{\o v})))$~) of sections of {\bbf the semisheaf of rings
$\phi _L(M_L(L_v))$\linebreak (resp. $\phi _R(M_R(L_{\o v}))$~),} also noted $\widetilde M_L(L_v)$ (resp. $\widetilde M_R(L_{\o v})$~).
\vskip 11pt

\item And  the set $\{\phi _R(M_{\o v_{\mu ,m_\mu }})\otimes \phi _L(M_{v_{\mu ,m_\mu }})\}_{\mu ,m_\mu }$ of differentiable bifunctions over the $\GL_2(L_{\o v}\times L_v)$-bisemimodule $M_R(L_{\o v})\otimes M_L(L_v)$ constitutes the set of bisections of {\bbf the bisemisheaf of rings $\phi _R(M_R(L_{\o v}))\otimes \phi _L(_L(L_v))$~.}\epr
\Ei

\section{Proposition}

{\em The bisemisheaf $\phi _R(M_R(L_{\o v}))\otimes \phi _L(M_L(L_v))$ of differentiable (bi)functions is a physical {\bfseries string field\/}.
}
\vskip 11pt

\bpr

\Bi
\item The bisections $\phi _R(M_{\o v_{\mu ,m_\mu }})\otimes \phi _L(M_{v_{\mu ,m_\mu }})$ of the bisemisheaf
$\phi _R(M_R(L_{\o v})) \otimes \phi _L(M_L(L_v))$ are $\CC$-valued differentiable bifunctions on the conjugacy class representatives
$(M_{\o v_{\mu ,m_\mu }}\otimes M_{v_{\mu ,m_\mu }})$ which are (tensor) products of two symmetric (closed) strings at $\mu $ quanta in such a way that
$(M_{\o v_{\mu ,m_\mu }}\otimes M_{v_{\mu ,m_\mu }})$ (and also 
$\phi _R(M_{\o v_{\mu ,m_\mu }})\otimes \phi _L(M_{v_{\mu ,m_\mu }})$~) behave like harmonic oscillators \cite{Pie1}.

\item So, the set $\{\phi _R(M_{\o v_{\mu ,m_\mu }})\otimes \phi _L(M_{v_{\mu ,m_\mu }})\}_{\mu ,m_\mu }$ of bisections of the bisemisheaf
$\phi _R(M_R(L_{\o v}))\linebreak \otimes \phi _L(M_L(L_v))$ is  a physical string field composed of packets of products of symmetric closed strings characterized by increasing numbers of quanta and behaving like harmonic oscillators.\epr
\Ei

\section{Getting a compact 4-dimensional semimanifold of space-time}

\Be
\item A smooth linear {\bf compact 3-dimensional semispace $M^I_\tau $ (resp. $M^I_{\o \tau }$~)}, restricted to the upper (resp. lower) half space, can be constructed by considering in the first place the compactification $C^r_L$ (resp. $C^r_R$~) of the conjugacy class representatives $M^I_{v_{\mu ,m_\mu }}$ (resp. $M^I_{\o v_{\mu ,m_\mu }}$~) of the ``space'' bilinear algebraic semigroup $\GL^I_2(L^I_{\o v}\times L^I_v)_r$ in the following way:

\Be
\item The real conjugacy clas representatives $\{M^I_{v_{\mu ,m_\mu }}\}_{m_\mu }$
(resp. $\{M^I_{\o v_{\mu ,m_\mu }}\}_{m_\mu }$~), $m_\mu $ varying, for each class $\mu $~, are compactified into a surface in such a way that they coincide with the complex equivalent
$M^I_{\omega _\mu }$ (resp. $M^I_{\o\omega _\mu }$~):
\[ \bigcup_c M^I_{v_{\mu ,m_\mu }}\subseteq M^I_{\omega _\mu }\qquad \text{(resp.} \quad
\bigcup_c M^I_{\o v_{\mu ,m_\mu }}\subseteq M^I_{\o\omega _\mu }\ )\]
as described in \cite{Pie3}, where $\bigcup\limits_c$ denotes the ``compact unions'' or compactification.

\item The set of these surfaces $\{M^I_{\omega _\mu }\}_\mu $ (resp. $\{M^I_{\o\omega _\mu }\}_\mu $~), $\mu $ varying, $1\le \mu \le q\le \infty $~, are also compactified into a 3-dimensional volume $M^I_{\tau _L}$ (resp. $M^I_{\tau _R}$~), which is assumed to be of ``space type'', in such a way that $M^I_{\tau _L}$ (resp. $M^I_{\tau _R}$~) be foliated by the surfaces $M^I_{\omega _\mu }$ (resp. $M^I_{\o\omega _\mu }$~).
\Ee

\item Consider then the complementary {\bbf ``time'' bilinear algebraic semigroup\linebreak  $\GL_2(L_{\o v}\times L_v)_t$\/} from which $\GL^I_2(L^I_{\o v}\times L^I_v)_r$ was generated according to proposition~2.7.

As for $\GL^I_2(L^I_{\o v}\times L^I_v)_r$~, the \lr linear conjugacy class representatives
$M_{v_{\mu ,m_\mu }}$ (resp. $M_{\o v_{\mu ,m_\mu }}$~) of the ``time'' bilinear algebraic semigroup $\GL^*_2(L^*_{\o v}\times L^*_v)_t$ are one-dimensional, corresponding then physically to ``time'' strings.

\item If the set $\{M_{v_{\mu ,m_\mu }}\}_{m_\mu,\mu  }$
(resp. $\{M_{\o v_{\mu ,m_\mu }}\}_{m_\mu,\mu  }$ of {\bbf the ``time'' conjugacy class representatives are glued together\/} under the map $C^{t-r}_L$ (resp. $C^{t-r}_R$~) {\bbf ``above'' the 3-dimensional volume $M^I_{\tau _L}$ (resp. $M^I_{\tau _R}$~) of space\/}, then we get a compactified 4-dimensional semispace of space-time $M^{T-S}_{\tau _L}$ (resp. $M^{T-S}_{\tau _R}$~).

\item {\bf In summary\/}, let:
\Bi
\item \bt[t]{ll}
& $C^r_L:\{M^I_{v_{\mu ,m_\mu }}\}_{\mu ,m_\mu }\to M^I_{\tau _L}=\bigcup\limits_C M^I_{\omega _\mu }$\\ (resp. & $C^r_R:\{M^I_{\o v_{\mu ,m_\mu }}\}_{\mu ,m_\mu }\to M^I_{\tau _R}=\bigcup\limits_C M^I_{\o\omega _\mu }$~)\te

 be the compactification of the one-dimensional conjugacy class representatives of space into a 3-dimensional semispace.

\item \bt[t]{ll}
& $C^{t-r}_L: M^I_{\tau _L}\cup \{M_{v_{\mu ,m_\mu }}\}_{\mu ,m_\mu } \to 
M^{T-S}_{\tau _L}=M^I_{\tau _L}\bigcup\limits_C\{M_{v_{\mu ,m_\mu }}\}_{\mu ,m_\mu }$\\
(resp. & $C^{t-r}_R: M^I_{\tau _R}\cup \{M_{\o v_{\mu ,m_\mu }}\}_{\mu ,m_\mu }\to M^{T-S}_{\tau _R}=M^I_{\tau _R}\bigcup\limits_C\{M_{\o v_{\mu ,m_\mu }}\}_{\mu ,m_\mu }$~)\te

 be the compactification of the one-dimensional conjugacy class representatives of time with the 3-dimensional compact semispace $M^I_{\tau _L}$ (resp. $M^I_{\tau _R}$~).
\Ei

\item A {\bbf 4-dimensional \lr semimanifold $\Ms^{T-S}_L$ (resp. $\Ms^{T-S}_R)$~)\/} is then defined from the 4-dimensional semispace $M^{T-S}_{\tau _L}$ (resp. $M^{T-S}_{\tau _R}$~) as a collection of charts $(M^{T-S}_{\tau _\mu },\phi _\mu )$ (resp. $(M^{T-S}_{\o\tau _\mu },\o\phi _\mu )$~), where:
\Bi
\item $M^{T-S}_{\tau _\mu }$ (resp. $M^{T-S}_{\o\tau _\mu }$~) are the compactified 4-dimensional semispaces obtained from the compactification of the linear conjugacy class representatives of $\GL_2(L_{\o v}\times L_v)_t$ and of $\GL^I_2(L^I_{\o v}\times L^I_v)_r$~.

\item $\phi _\mu $ (resp. $\o\phi _\mu $~) are one-to-one maps of $M^{T-S}_{\tau _\mu }$ (resp. $M^{T-S}_{\o\tau _\mu }$~) to open sets in the upper (resp. lower) half of $\rit^4$~,
\Ei

such that
\Be
\item $\Ms^{T-S}_L=\bigcup_\mu M^{T-S}_{\tau _\mu }$ \quad (resp. $\Ms^{T-S}_R=\bigcup_\mu M^{T-S}_{\o\tau _\mu }$~).
\item if \quad $M^{T-S}_{\tau _\mu }\cap M^{T-S}_{\tau _\nu }$ \quad (resp. \quad $M^{T-S}_{\o\tau _\mu }\cap M^{T-S}_{\o\tau _\nu }$~) 
\quad
is non-empty, then the map
\begin{alignat*}{3}
\phi _\mu \circ \phi _\nu ^{-1}&: \quad \phi _\nu (M^{T-S}_{\tau _\mu }\cap M^{T-S}_{\tau _\nu })
&\To& \phi _\mu (M^{T-S}_{\tau _\mu }\cap M^{T-S}_{\tau _\nu })\\
\text{(resp.} \quad 
\o\phi _\mu \circ \o\phi _\nu ^{-1}&: \quad \o\phi _\nu (M^{T-S}_{\o\tau _\mu }\cap M^{T-S}_{\o\tau _\nu })
&\To& \o\phi _\mu (M^{T-S}_{\o\tau _\mu }\cap M^{T-S}_{\o\tau _\nu })\; )\end{alignat*}
is a map of open upper (resp. lower) half subset of $\rit^4$ to an open (resp. lower) half subset of $\rit^4$~.
\Ee

\item {\bf The composition of maps\/}:
\begin{alignat*}{3}
C^{t-r}_L\circ C^r_L&: \quad
\{M_{v_{\mu ,m_\mu }}\}_{\mu ,m_\mu }\cup \{M^I_{v_{\mu ,m_\mu }}\}_{\mu ,m_\mu }&\To& M^{T-S}_{\tau _L}\\
\text{(resp.} \quad
C^{t-r}_R\circ C^r_R&: \quad
\{M_{\o v_{\mu ,m_\mu }}\}_{\mu ,m_\mu }\cup \{M^I_{\o v_{\mu ,m_\mu }}\}_{\mu ,m_\mu }&\To& M^{T-S}_{\tau _R}\; ),
\end{alignat*}
compactifying one-dimensional conjugacy class representatives of space with one-dimensional conjugacy class representatives of time into a compact 4-dimensional upper (resp. lower) semispace $M^{T-S}_{\tau _L}$ (resp. $M^{T-S}_{\tau _R}$~) of space-time, {\bbf allows to get a classical continuous (Riemann) 4-dimensional semispace from locally compact one-dimensional conjugacy class representatives of space-time at the Planck scale, the compactification map $C^{t-r}_L\circ C^r_L$ (resp. $C^{t-r}_R\circ C^r_R$~) corresponding to a change of scale\/}.
\Ee

\section{Proposition}

{\em The product $(C^{t-r}_R\circ C^r_R) \times(C^{t-r}_L\circ C^r_L)$ of the 4-dimensional compactifications introduced in section 2.10.6. sends the direct sum of the representation space $\Repsp (\GL_2^*(L^*_{\o v}\times L_v^*)_t)$ of the reduced ``time'' bilinear algebraic semigroup $\GL_2^*(L^*_{\o v}\times L_v^*)_t$ and of the representation space $\Repsp
(\GL_2^I(L^I_{\o v}\times L_v^I)_r)$ of the complementary ``space'' bilinear algebraic semigroup $\GL_2^I(L^I_{\o v}\times L_v^I)$ into the product $(\bM^{T-S}_{\tau _R} \otimes
\bM^{T-S}_{\tau _L})$ of the 4-dimensional compact lower semispace
$\bM^{T-S}_{\tau _R}$ by its upper equivalent $\bM^{T-S}_{\tau _L}$ according to:
\begin{multline*}
(C^{t-r}_R\circ C^r_R) \times (C^{t-r}_L\circ C^r_L) : \quad
\Repsp (\GL_2^* ( L^*_{\o v}\times L_v^*)_t) \oplus
\Repsp (\GL_2^I(L^I_{\o v}\times L_v^I))_r) \\
\To (\bM^{T-S}_{\tau _R} \otimes
\bM^{T-S}_{\tau _L})\end{multline*}
in such a way that:
\Bi
\item each point of $(\bM^{T-S}_{\tau _R} \otimes
\bM^{T-S}_{\tau _L})$ is in fact a bipoint characterized by a metric tensor $g$~.
\item each bipoint of $(\bM^{T-S}_{\tau _R} \otimes
\bM^{T-S}_{\tau _L})$ is in one-to-one correspondence with a bipoint of
$\Repsp (\GL_2^* ( L^*_{\o v}\times L_v^*)_t) \oplus
\Repsp (\GL_2^I(L^I_{\o v}\times L_v^I))_r)$ so that
$(C^{t-r}_R\circ C^r_R) \times (C^{t-r}_L\circ C^r_L)$ is an isomorphism.
\Ei
}
\vskip 11pt

\bpr
\Be\item
the fact that $(C^{t-r}_R\circ C^r_R) \times (C^{t-r}_L\circ C^r_L)$ is a (bi)isomorphism results from its compactifying nature introduced in section 2.10.

\item Each point $\Ps$ of $(\bM^{T-S}_{\tau _R} \otimes
\bM^{T-S}_{\tau _L})$ is a bipoint $\Ps_R\times \Ps_L$~, product of a point $\Ps_L\in M^{T-S}_{\tau _L}$ and of a point $\Ps_R\in M^{T-S}_{\tau _R}$~.

A metric tensor $g$ at $\Ps\equiv \Ps_R\times \Ps_L$ has for components $g^b_a = g(E_a,E^b)$ (or $g_{a_b}$~) with respect to basis vectors $E_a\in  \bM^{T-S}_{\tau _R}$and $E_b\in  \bM^{T-S}_{\tau _L}$~, $a=t,x,y,z$~, $b=t,x,y,z$~, in such a way that $g (E_a,E^b)$ is a scalar product between basis vectors.

\item The compactified bisemispace $ \bM^{T-S}_{\tau _R} \otimes  \bM^{T-S}_{\tau _L}$ can be understood if it is realized that each bipoint $(\Ps_R\times \Ps_L)$ of it is in one-to-one correspondence with a bipoint $(p_R\times p_L)$ belonging to the product, right by left, of two {\bf one-dimensional\/} symmetric conjugacy class representatives of
$\Repsp (\GL_2^* ( L^*_{\o v}\times L_v^*)_t) $ or of $
\Repsp (\GL_2^I(L^I_{\o v}\times L_v^I))_r)$~, if it is referred to section 2.10.\epr
\Ee

%% file: CPTrack3.tex
\chapter{The equations of the internal dynamics of a bisemifermion}

\thispagestyle{empty}

\section{Singularizations on the internal vacuum semi\-sheaves}

\Bi
\item It was seen in proposition 2.7 that {\bf the internal vacuum space-time structure\/} of an elementary (bisemi)fermion is given by the direct sum of the representation spaces of the bilinear algebraic semigroups $\GL_2(L_{\o v}\times L_v)_t$ and $\GL^I_2(L^I_{\o v}\times L^I_v)_r$
according to:
\[ (M^T_{ST_R}\otimes M^T_{ST_L}) \oplus (M^S_{ST_R}\otimes M^S_{ST_L})
= \Repsp (\GL_2 (L_{\o v}\times L_v)_t)\oplus \Repsp (\GL^I_2 (L^I_{\o v}\times L^I_v)_r)\]
where:
\Bi
\item $M^T_{ST_L}\equiv \Repsp(T_2 (L_{\o v}\times L_v)_t)$ \quad (resp. 
$M^T_{ST_R}\equiv \Repsp(T_2^t(L_{\o v})_t)$~) 

represents the algebraic time structure of the internal vacuum of a \lr semifermion.
\item $M^S_{ST_L}\equiv \Repsp(T^I_2(L^I_v))$ \quad (resp. 
$M^S_{ST_R}\equiv \Repsp(T_2^{t,I}(L^I_{\o v}))$~) 

represents the algebraic space structure of the internal vacuum of a \lr semifermion.
\Ei
\vskip 11pt

\item {\bf The semisheaf of differentiable functions\/} on $M^T_{ST_L}$ (resp. $M^T_{ST_R}$~) and on $M^S_{ST_L}$ (resp. $M^S_{ST_R}$~) is noted respectively 
$\widetilde M^T_{ST_L}$ (resp. $\widetilde M^T_{ST_R}$~) and $\widetilde M^S_{ST_L}$ (resp. $\widetilde M^S_{ST_R}$~).

The bisemisheaves $(\widetilde M^T_{ST_L}\otimes\widetilde M^T_{ST_R})$ and
$(\widetilde M^S_{ST_L}\otimes \widetilde M^S_{ST_R})$ are, respectively, a ``time'' and a ``space'' string field of the internal vacuum of an elementary bisemifermion.
\vskip 11pt

\item As these internal vacuum string fields have a spatial extension of the order of the Planck length, they are submitted to strong fluctuations generating {\bf degenerate singularities\/} on the sections (or strings) of these, as described mathematically in \cite{Pie4}.

Taking into account that the time (and space) left and right semisheaves are symmetrical by construction and localized in small open balls, it is reasonable to assume that the singularities are generated symmetrically on the corresponding sections respectively in the upper and lower half spaces.
\vskip 11pt

\item By this way, it is easy to understand that {\bf the metric tensor\/} $g^b_a$ at each bipoint $(\Ps_R\times \Ps_L)$ of a bisection 
$(\widetilde M^T_{\o v_{\mu ,m_\mu }} \otimes \widetilde M^T_{v_{\mu ,m_\mu }})$ (also noted
$\phi _R(M_{\o v_{\mu ,m_\mu }})\otimes \phi _L(M_{v_{\mu ,m_\mu }})$ in proposition 2.9) of
$(\widetilde M^T_{ST_R}\otimes\widetilde M^T_{ST_L})$ or of a bisection 
$(\widetilde M^S_{\o v_{\nu ,m_\nu }} \otimes \widetilde M^S_{v_{\nu ,m_\nu }})$ of
$(\widetilde M^S_{ST_R}\otimes\widetilde M^S_{ST_L})$ is not constant but varies according to the singularities on the sections.
\Ei
\vskip 11pt

\section{Versal deformation and its blowup}

\Bi
\item Under a strong external perturbation, {\bf a degenerate singularity\/} of multiplicity inferior or equal to 3 is assumed to be generated on each section $\widetilde M^S_{v_{\nu ,m_\nu }}$
(resp. $\widetilde M^S_{\o v_{\nu ,m_\nu }}$~) of the ``space'' semisheaf $\widetilde M^S_{ST_L}$ (resp. $\widetilde M^S_{ST_R}$~).
\vskip 11pt

\item Then, {\bf a versal deformation\/} of $\widetilde M^S_{ST_L}$ (resp. $\widetilde M^S_{ST_R}$~) can be envisaged as given by the fibre bundle:
\begin{alignat*}{3}
D_{S_L}: \quad \widetilde M^S_{ST_L} \times \theta _{S_L} &\To& \widetilde M^S_{ST_L}\\
\text{(resp.} \quad 
D_{S_R}: \quad \widetilde M^S_{ST_R} \times \theta _{S_R} &\To& \widetilde M^S_{ST_R}\ )
\end{alignat*}
where the fibre $\theta _{S_L}=\{\theta _1(\omega ^1_L),\theta _2(\omega^2_L),\theta _3(\omega ^3_L)\}$ (resp. $\theta _{S_R}=\{\theta _1(\omega ^1_R),\theta _2(\omega^2_R),\theta _3(\omega ^3_R)\}$~) is composed of the set of three sheaves of the base $S_L$ (resp. $S_R$~) of the versal deformation in such a way that the $\omega ^i_L$ (resp. $\omega ^i_R$~, $1\le i\le 3$~, are the generators of the base of the quotient algebra.
\vskip 11pt

\item We refer to \cite{Pie4} for more complete developments of the versal deformation and of its blowup succinctly recalled in the following.
\vskip 11pt

\item {\bf The blowup of the versal deformation\/} allows to generate a semisheaf
$\widetilde M^S_{MG_L}$ (resp. $\widetilde M^S_{MG_R}$~) covering the space internal vacuum semisheaf $\widetilde M^S_{ST_L}$ (resp. $\widetilde M^S_{ST_R}$~) by means of the spreading-out isomorphism:
\[ SO T_L = \tau _{v_{\omega _L}}\circ \pi _{s_L} \qquad \text{(resp.} \quad
SO T_R = \tau _{v_{\omega _R}}\circ \pi _{s_R}\ )\]
where:
\Bi
\item \bt[t]{rl}
& $\pi _{s_L} : \quad  \widetilde M^S_{ST_L}\times \theta _{S_L} \To
\widetilde M^S_{ST_L}\oplus \theta _{S_L} $\\
(resp. \quad & $\pi _{s_R} : \quad  \widetilde M^S_{ST_R}\times \theta _{S_R} \To
\widetilde M^S_{ST_R}\oplus \theta _{S_R} $~)\te

is an endomorphism, based on a Galois antiautomorphism \cite{Pie5}, which disconnects the fibre $\theta _{S_L}$ (resp. $\theta _{S_R}$~) from $\widetilde M^S_{ST_L}$ (resp.  $ M^S_{ST_R}$~).
\item $\tau _{v_{\omega _L}}$ (resp. $\tau _{v_{\omega _R}}$~) is the projective map:
\[ \tau _{v_{\omega _L}}: \quad \TAN(\theta _{S_L})\To \theta _{S_L} \qquad \text{(resp.} \quad
\tau _{v_{\omega _R}}: \quad \TAN(\theta _{S_R})\To \theta _{S_R}\ )\]
of the vertical tangent bundle $T_{v_{\omega _L}}$ (resp. $T_{v_{\omega _R}}$~) sending $\theta _{S_L}$ (resp. $\theta _{S_R}$~) into the total tangent space $\TAN(\theta _{S_L})$ (resp. $\TAN(\theta _{S_R})$~).
\Ei

By this way, the three functions $\omega ^i_L(v_{\nu ,m_\nu })$ (resp. $\omega ^i_R(\o v_{\nu ,m_\nu })$~) of the base of the versal deformation are projected above each section 
$\widetilde M^S_{v_{\nu ,m_\nu }}$ (resp. $\widetilde M^S_{\o v_{\nu ,m_\nu }}$~) in the vertical tangent space.

Being glued together, these three functions generate the sections
$\widetilde M^{S_p}_{MG_{v_{\nu ,m_\nu }}}$ (resp. $\widetilde M^{S_p}_{MG_{\o v_{\nu ,m_\nu }}}$~) of a semisheaf $\widetilde M^{S_p}_{MG_L}$ (resp. $\widetilde M^{S_p}_{MG_R}$~) (called middle ground) which covers the internal vacuum semisheaf.

If the numbers of quanta on the sections
$\widetilde M^{S_p}_{MG_{v_{\nu ,m_\nu }}}$ (resp. $\widetilde M^{S_p}_{MG_{\o v_{\nu ,m_\nu }}}$~) of the {\bbf middle ground semisheaf 
$\widetilde M^{S_p}_{MG_L}$ (resp. $\widetilde M^{S_p}_{MG_R}$~)\/} are equal to the numbers of quanta on the sections 
$\widetilde M^{S}_{v_{\nu ,m_\nu }}$ (resp. $\widetilde M^{S}_{\o v_{\nu ,m_\nu }}$~)
(rewritten according to $\widetilde M^{S}_{ST_{v_{\nu ,m_\nu }}}$ (resp. $\widetilde M^{S}_{ST_{\o v_{\nu ,m_\nu }}}$~)) of the internal vacuum semisheaf
$\widetilde M^{S}_{ST_L}$ (resp. $\widetilde M^{S}_{ST_R}$~),then these sections 
$\widetilde M^{S_p}_{MG_{v_{\nu ,m_\nu }}}$ (resp. $\widetilde M^{S_p}_{MG_{\o v_{\nu ,m_\nu }}}$~)
 are open strings covering the closed strings
$\widetilde M^{S}_{ST_{v_{\nu ,m_\nu }}}$ (resp. $\widetilde M^{S}_{ST_{\o v_{\nu ,m_\nu }}}$~) of $\widetilde M^{S}_{ST_L}$ (resp. $\widetilde M^{S}_{ST_R}$~).
\Ei
\vskip 11pt

\section{Proposition}

{\em The inverse of the projective map $\tau _{v_{\omega _L}}$ (resp. $\tau _{v_{\omega _R}}$~) of the tangent bundle $T _{v_{\omega _L}}$ (resp. $T _{v_{\omega _R}}$~) of the spreading-out isomorphism is the elliptic operator:
\begin{align*}
DT^S_{L;MG} &= \L\{ i\ \F{\hbar_{MG}}{c_{t\to r;MG}}\ \F\partial{\partial x},
i\ \F{\hbar_{MG}}{c_{t\to r;MG}}\ \F\partial{\partial y},
i\ \F{\hbar_{MG}}{c_{t\to r;MG}}\ \F\partial{\partial z}\R\}\\[11pt]
\biggl(\text{resp.} \quad 
DT^S_{R;MG} &= \L\{ -i\ \F{\hbar_{MG}}{c_{t\to r;MG}}\ \F\partial{\partial x},
-i\ \F{\hbar_{MG}}{c_{t\to r;MG}}\ \F\partial{\partial y},
-i\ \F{\hbar_{MG}}{c_{t\to r;MG}}\ \F\partial{\partial z}\R\}\ \biggr),\end{align*}
where the constants $\hbar_{MG}$ and $c_{t\to r;MG}$ are defined in chapter 3 of \cite{Pie5}, sending the semisheaf $\widetilde M^{S}_{MG_L}$ (resp. $\widetilde M^{S}_{MG_R}$~) into the perverse semisheaf $\widetilde M^{S_p}_{MG_L}$ (resp. $\widetilde M^{S_p}_{MG_R}$~) according to:
\[DT^S_{L;MG}: \quad \widetilde M^S_{MG_L} \To \widetilde M^{S_p}_{MG_L} \qquad
\text{(resp.} \quad 
DT^S_{R;MG}: \quad \widetilde M^S_{MG_R} \To \widetilde M^{S_p}_{MG_R}\ ).\]
}
\vskip 11pt

\bpr The elliptic operator $DT^S_{L;MG}$ (resp. $DT^S_{R;MG}$~) maps the semisheaf 
$\widetilde M^S_{MG_L}$ (resp. $\widetilde M^S_{MG_R}$~) into its perverse equivalent 
$\widetilde M^{S_p}_{MG_L}$ (resp. $\widetilde M^{S_p}_{MG_R}$~) since this latter belongs to the derived category of string semifields shifted in the three geometrical dimensions of space, a string semifield being given by a semisheaf, for example 
$\widetilde M^S_{MG_L}$ (resp. $\widetilde M^S_{MG_R}$~).\epr
\vskip 11pt

\section[Generation of perverse mass semisheaves $\widetilde M^{S_p}_{M_L}$ and 
$\widetilde M^{S_p}_{M_R}$]{\bbf Generation of perverse mass semisheaves $\widetilde M^{S_p}_{M_L}$ and 
$\widetilde M^{S_p}_{M_R}$}

\Bi
\item As the singularities on the sections of the internal vacuum semisheaf $\widetilde M^{S}_{ST_L}$ (resp. 
$\widetilde M^{S}_{ST_R}$~) have a multiplicity inferior or equal to 3~, the functions $\omega ^i_L(v_{\nu ,m_\nu })$ (resp. $\omega ^i_r(\o v_{\nu ,m_\nu })$~) of the quotient algebra of the versal deformation of $\widetilde M^{S}_{ST_L}$ (resp. 
$\widetilde M^{S}_{ST_R}$~) can  have degenerate singularities of multiplicity one.
\vskip 11pt

\item So, the middle ground semisheaf $\widetilde M^{S}_{MG_L}$ (resp. 
$\widetilde M^{S}_{MG_R}$~), of which sections $\widetilde M^{S}_{MG_{v_{\nu ,m_\nu }}}$ (resp. 
$\widetilde M^{S}_{MG_{\o v_{\nu ,m_\nu }}}$~) are the functions $\omega ^i_L(v_{\nu ,m_\nu })$ (resp. $\omega ^i_R(\o v_{\nu ,m_\nu })$~), $1\le i\le 3$~, glued together, can undergo a versal deformation and a blowup of it according to:
\begin{alignat*}{3}
SO T^{(MG)}_L\circ D^{(MG)}_{S_L} &: \quad 
\widetilde M^{S}_{MG_L}&\To& \widetilde M^{S}_{MG_L}\oplus \widetilde M^{S_p}_{M_L}\\
\text{(resp.} \quad 
SO T^{(MG)}_R\circ D^{(MG)}_{S_R} &: \quad 
\widetilde M^{S}_{MG_R}&\To&\widetilde M^{S}_{MG_R}\oplus \widetilde M^{S_p}_{M_R}\ )
\end{alignat*}
where:
\Bi
\item \bt[t]{ll}
& $ D^{(MG)}_{S_L} : 
\widetilde M^{S}_{MG_L}\times \theta ^{(MG)}_{S_L}\to \widetilde M^{S}_{MG_L}$ \\
(resp. &
$ D^{(MG)}_{S_R} : 
\widetilde M^{S}_{MG_R}\times \theta ^{(MG)}_{S_R}\to \widetilde M^{S}_{MG_R}$~)\te

 is the versal deformation of
$\widetilde M^{S}_{MG_L}$ (resp. $ \widetilde M^{S}_{MG_R}$~).

\item \bt[t]{ll}
& $ SO T^{(MG)}_{L} : 
\widetilde M^{S}_{MG_L}\times \theta ^{(MG)}_{S_L}\to \widetilde M^{S}_{MG_L} \oplus \widetilde M^{S_p}_{M_L}$ \\
(resp. &
$ SO T^{(MG)}_{R} : 
\widetilde M^{S}_{MG_R}\times \theta ^{(MG)}_{S_R}\to \widetilde M^{S}_{MG_R} \oplus \widetilde M^{S_p}_{M_R}$~)\te

 in such a way that
$\widetilde M^{S_p}_{M_L}$ (resp. $ \widetilde M^{S_p}_{M_R}$~) is the perverse mass semisheaf generated from
$\widetilde M^{S}_{M_L}\equiv \theta ^{(MG)}_{S_L}$ (resp. $\widetilde M^{S}_{M_R}\equiv \theta ^{(MG)}_{S_R}$~) by the map:
\[ DT^S_{L;M} : \quad \widetilde M^{S}_{M_L}\To \widetilde M ^{S_p}_{M_L} \qquad \text{(resp.} \quad 
DT^S_{R;M} : \quad \widetilde M^{S}_{M_R}\To \widetilde M ^{S_p}_{M_R}\ )\]
so that
\begin{align*}
DT^S_{L;M} &=\L\{ i\ \F\hbar{c_{t\to r;M}}\ \F\partial{\partial x} ,
i\ \F\hbar{c_{t\to r;M}}\ \F\partial{\partial y} ,
i\ \F\hbar{c_{t\to r;M}}\ \F\partial{\partial z} \R\}\\[11pt]
\Biggl(\text{resp.} \quad 
DT^S_{R;M} &=\L\{ -i\ \F\hbar{c_{t\to r;M}}\ \F\partial{\partial x} ,
-i\ \F\hbar{c_{t\to r;M}}\ \F\partial{\partial y} ,
-i\ \F\hbar{c_{t\to r;M}}\ \F\partial{\partial z} \R\}\ \Biggr)\end{align*}
corresponds to the inverse of the projective map
\[ \tau ^{(M)}_{v_{\omega _L}}: \quad \TAN(\theta ^{(MG)}_{S_L}) \To \theta ^{(MG)}_{S_L}
\qquad \text{(resp.}\quad 
\tau ^{(M)}_{v_{\omega _R}}: \quad \TAN(\theta ^{(MG)}_{S_R}) \To \theta ^{(MG)}_{S_R}\ )\]
of the vertical tangent bundle of the blowup of the versal deformation of the middle ground semisheaf $\widetilde M^{S}_{MG_L}$ (resp. $\widetilde M^{S}_{MG_R}$~).
\Ei
\vskip 11pt

\item {\bf The sections $\widetilde M^{S_p}_{M_{v_{\nu ,m_\nu }}}$ (resp. 
$\widetilde M^{S_p}_{M_{\o v_{\nu ,m_\nu }}}$~) of the perverse ``mass'' semisheaf
$\widetilde M^{S_p}_{M_L}$ (resp. $\widetilde M^{S_p}_{M_R}$~) cover the corresponding sections of the ``middle ground'' and ``internal vacuum'' semisheaves\/}
$\widetilde M^{S_p}_{MG_L}$ (resp. $\widetilde M^{S_p}_{MG_L}$~) and
$\widetilde M^{S}_{ST_L}$ (resp. $\widetilde M^{S}_{ST_R}$~): they are open strings if the numbers of quanta on their sections are inferior or equal to the number of quanta on the sections of the ``middle ground'' semisheaves
$\widetilde M^{S_p}_{MG_L}$ (resp. $\widetilde M^{S_p}_{MG_R}$~).
\Ei
\vskip 11pt

\section{Embedding of ``internal vacuum'', ``middle ground'' and ``mass'' semisheaves of space}

\Bi
\item So, the ``middle ground'' and ``mass'' semisheaves of space
$\widetilde M^{S}_{MG_L}$ (resp. $\widetilde M^{S}_{MG_R}$~) and
$\widetilde M^{S}_{M_L}$ (resp. $\widetilde M^{S}_{M_R}$~) can be generated by versal deformations and blowups of these from the ``internal vacuum'' semisheaf
$\widetilde M^{S}_{ST_L}$ (resp. $\widetilde M^{S}_{ST_R}$~) of space leading to the {\bf embedding\/}:
\[ \widetilde M^{S}_{ST_L} \subset \widetilde M^{S}_{MG_L} \subset \widetilde M^{S}_{M_L}\qquad \text{(resp.} \quad 
\widetilde M^{S}_{ST_R} \subset \widetilde M^{S}_{MG_R} \subset \widetilde M^{S}_{M_R}\ ).\]

\item {\bf The perverse ``middle ground' and ``mass'' semisheaves 
$\widetilde M^{S_p}_{MG_L} $ (resp. $\widetilde M^{S_p}_{MG_R} $~) and
$\widetilde M^{S_p}_{M_L} $ (resp. $\widetilde M^{S_p}_{M_R} $~)\/}, belonging to the derived category of string semifields, {\bf are of contracting nature\/}, while the {\bf ``internal vacuum'' semisheaf of space 
$\widetilde M^{S}_{ST_L} $ (resp. $\widetilde M^{S}_{ST_R} $~) is of expanding nature\/}: so, it was assumed in \cite{Pie5} that the $3D$-differential operator
\begin{align*}
T^S_{L;ST} &= \L\{ i\ \F{\hbar_{ST}}{c_{t\to r;ST}}\ dx ,
i\ \F{\hbar_{ST}}{c_{t\to r;ST}}\ dy ,
i\ \F{\hbar_{ST}}{c_{t\to r;ST}}\ dz\R\}\\[11pt]
\Biggl(\text{resp.} \quad 
T^S_{R;ST} &= \L\{ -i\ \F{\hbar_{ST}}{c_{t\to r;ST}}\ dx ,
-i\ \F{\hbar_{ST}}{c_{t\to r;ST}}\ dy ,
-i\ \F{\hbar_{ST}}{c_{t\to r;ST}}\ dz\R\}\ \Biggr)\end{align*}
applies to all the sections of
$\widetilde M^{S}_{ST_L} $ (resp. $\widetilde M^{S}_{ST_R} $~) according to:
\[ T^S_{L;ST}: \quad \widetilde M^{S}_{ST_L} \To \widetilde M^{S_p}_{ST_L} \qquad \text{(resp.} \quad
T^S_{R;ST}: \quad \widetilde M^{S}_{ST_R} \To \widetilde M^{S_p}_{ST_R} \ )\]
where $\widetilde M^{S_p}_{ST_L} $ (resp. $\widetilde M^{S_p}_{ST_R} $~) was written abusively as a perverse semisheaf.

Similarly, we have the following embedding
\[ \widetilde M^{S_p}_{ST_L} \subset \widetilde M^{S_p}_{MG_L} \subset \widetilde M^{S_p}_{M_L}\qquad \text{(resp.} \quad 
\widetilde M^{S_p}_{ST_R} \subset \widetilde M^{S_p}_{MG_R} \subset \widetilde M^{S_p}_{M_R}\ )\]
between ``perverse'' semisheaves.
\Ei
\vskip 11pt

\section{Middle ground and mass semisheaves of time}

\Bi
\item Similarly, as the ``middle ground'' and ``mass'' semisheaves of space are generated by versal deformations and blowups fom the ``internal vacuum'' semisheaf of space 
$\widetilde M^{S}_{ST_L}$ (resp. $\widetilde M^{S}_{ST_R}$~), the ``middle ground'' and ``mass'' semisheaves of time
$\widetilde M^{T}_{MG_L}$ (resp. $\widetilde M^{T}_{MG_R}$~) and
$\widetilde M^{T}_{M_L}$ (resp. $\widetilde M^{T}_{M_R}$~) can be generated:
\Bi
\item either by versal deformations and blowups from the ``internal vaccuum'' semisheaf of time
$\widetilde M^{T}_{ST_L}$ (resp. $\widetilde M^{T}_{ST_R}$~).
\item or from the respective semisheaves of space by the composition of maps:
\begin{alignat*}{3}
\gamma _{r_L\to t_L}^{(MG)} \circ E^{(MG)}_L&: \quad 
\widetilde M^{S}_{MG_L}&\To&\widetilde M^{T}_{MG_L}\\ 
\text{and} \quad
\gamma _{r_L\to t_L}^{(M)} \circ E^{(M)}_L&: \quad 
\widetilde M^{S}_{M_L}&\To&\widetilde M^{T}_{M_L}\\[11pt]
\text{(resp.} \quad
\gamma _{r_R\to t_R}^{(MG)} \circ E^{(MG)}_R&: \quad 
\widetilde M^{S}_{MG_R}&\To&\widetilde M^{T}_{MG_R}\\
\text{and} \quad
\gamma _{r_R\to t_R}^{(M)} \circ E^{(M)}_R&: \quad 
\widetilde M^{S}_{M_R}&\To&\widetilde M^{T}_{M_R}\ ),\end{alignat*}
as described in proposition 2.7, where $E^{(MG)}_L$ and $E^{(M)}_L$ (resp. $E^{(MG)}_R$ and $E^{(M)}_R$~) are endomorphisms based on Galois antiautomorphisms.
\Ei

As for the semisheaves of space, we have {\bf for the semisheaves of time the embedding\/}:
\[ \widetilde M^{T}_{ST_L} \subset \widetilde M^{T}_{MG_L} \subset \widetilde M^{T}_{M_L}\qquad \text{(resp.} \quad 
\widetilde M^{T}_{ST_R} \subset \widetilde M^{T}_{MG_R} \subset \widetilde M^{T}_{M_R}\ )\]
{\bf as well as for the perverse semisheaves of time\/}:
\[ \widetilde M^{T_p}_{ST_L} \subset \widetilde M^{T_p}_{MG_L} \subset \widetilde M^{T_p}_{M_L}\qquad \text{(resp.} \quad 
\widetilde M^{T_p}_{ST_R} \subset \widetilde M^{T_p}_{MG_R} \subset \widetilde M^{T_p}_{M_R}\ ).\]
These perverse semisheaves of time result form the morphisms:
\begin{alignat*}{3}
 T^T_{L;ST}&: \quad \widetilde M^{T}_{ST_L} &\To & \widetilde M^{T_p}_{ST_L}\;, \\
DT^T_{L;MG}&: \quad \widetilde M^{T}_{MG_L} &\To & \widetilde M^{T_p}_{MG_L}\;, \\
DT^T_{L;M}&: \quad \widetilde M^{T}_{M_L} &\To& \widetilde M^{T_p}_{M_L}\;, \end{alignat*}
where
\[ T^T_{L;ST}=i\ \F{\hbar_{ST}}{c_{t\to r;ST}}\ dt_0\;, \quad
DT^T_{L;MG}=i\ \F{\hbar_{MG}}{c_{t\to r;MG}}\ \F\partial{\partial t_0}\;, \quad
DT^T_{L;M}=i\ \F{\hbar_{M}}{c_{t\to r;M}}\ \F\partial{\partial t_0}\]
(the right cases are handled similarly).
\Ei
\vskip 11pt

\section{Proposition}

{\em Let $(\widetilde M^{T-S}_{ST_R}\otimes \widetilde M^{T-S}_{ST_L})$ be the ``internal vacuum'' bisemisheaf or string field of an elementary bisemifermion.

Then, by versal deformations and blowups of these, {\bfseries the  ``internal vacuum'' string field can generate the two covering ``middle ground'' and ``mass'' string fields of space-time\/}
\[ (\widetilde M^{T-S}_{MG_R}\otimes \widetilde M^{T-S}_{MG_L})
\qquad \and \qquad 
(\widetilde M^{T-S}_{M_R}\otimes \widetilde M^{T-S}_{M_L})\]
leading to the embedding:
\[ (\widetilde M^{T-S}_{ST_R}\otimes \widetilde M^{T-S}_{ST_L})
\subset (\widetilde M^{T-S}_{MG_R}\otimes \widetilde M^{T-S}_{MG_L})
\subset (\widetilde M^{T-S}_{M_R}\otimes \widetilde M^{T-S}_{M_L}).\]
}
\vskip 11pt

\bpr
\Bi
\item According to section 3.1, the ``internal vacuum'' string field is given by:
\[ (\widetilde M^{T}_{ST_R}\otimes \widetilde M^{T}_{ST_L})\oplus
(\widetilde M^{S}_{ST_R}\otimes \widetilde M^{S}_{ST_L})\]
where $(\widetilde M^{T}_{ST_R}\otimes \widetilde M^{T}_{ST_L})$ (resp.
$(\widetilde M^{S}_{ST_R}\otimes \widetilde M^{S}_{ST_L})$~) is the time (resp. space) string field.

But, this ``internal vacuum'' string field of space time is not complete, because it does not allow interactions between the time string field and the space string field.

\item A more general approach would consider the following ``internal vacuum'' string field:
\begin{multline*}
(\widetilde M^{T-S}_{ST_R}\otimes \widetilde M^{T-S}_{ST_L})
\equiv (\widetilde M^{T}_{ST_R}\oplus \widetilde M^{S}_{ST_R})
\otimes (\widetilde M^{T}_{ST_L}\oplus \widetilde M^{S}_{ST_L})\\
= (\widetilde M^{T}_{ST_R}\otimes \widetilde M^{T}_{ST_L})
\oplus (\widetilde M^{S}_{ST_R}\otimes \widetilde M^{S}_{ST_L})
\oplus (\widetilde M^{T}_{ST_R}\otimes \widetilde M^{S}_{ST_L})
\oplus (\widetilde M^{S}_{ST_R}\otimes \widetilde M^{T}_{ST_L})
\end{multline*}
constituting the {\bf completely reducible non orthogonal bilinear representation space in bisemisheaves\/} \cite{Pie4}
of $\GL_{2(T+S)}(L_{\o v}\times L_v))$ according to:
\begin{multline*}
\Repsp \theta (\GL_{2(T+S)}(L_{\o v}\times L_v))\\
= \Repsp \theta (\GL_{2(T)}(L_{\o v}\times L_v))
\oplus \Repsp \theta (\GL_{2(S)}(L^I_{\o v}\times L^I_v))\qquad\\
\oplus \Repsp \theta (T^t_{2(T)}(L_{\o v})\times T_{2(S)}(L^I_v))
\oplus \Repsp \theta (T^t_{2(S)}(L^I_{\o v})\times T_{2(T)}(L_v))
\end{multline*}
where
\begin{align*}
(\widetilde M^{T}_{ST_R}\otimes \widetilde M^{S}_{ST_L})
\equiv \Repsp \theta (T^t_{2(T)}(L_{\o v})\times T_{2(S)}(L^I_v))\\
\and \quad 
(\widetilde M^{S}_{ST_R}\otimes \widetilde M^{T}_{ST_L})
\equiv \Repsp \theta (T^t_{2(S)}(L^I_{\o v})\times T_{2(T)}(L_v))\end{align*}
are mixed bisemisheaves over off diagonal bilinear representation spaces of ``time-space'' and ``space-time'' responsible for the electric charge of the considered (bisemi)- fermion as developed in \cite{Pie5}.

\item Referring to the preceding sections, the ``internal vacuum'' bisemisheaf
$(\widetilde M^{T-S}_{ST_R}\otimes \widetilde M^{T-S}_{ST_L})$ generates by versal deformations and blowups of these the ``middle ground'' and ``mass'' bisemisheaves
$(\widetilde M^{T-S}_{MG_R}\otimes \widetilde M^{T-S}_{MG_L})$ and
$(\widetilde M^{T-S}_{M_R}\otimes \widetilde M^{T-S}_{M_L})$ which are embedded as announced in this proposition.\epr
\Ei
\vskip 11pt

\section{Proposition}

{\em \Be
\item The corresponding perverse bisemisheaves (or fields)
$(\widetilde M^{T_p-S_p}_{ST_R}\otimes \widetilde M^{T_p-S_p}_{ST_L})$~,
$(\widetilde M^{T_p-S_p}_{MG_R}\otimes \widetilde M^{T_p-S_p}_{MG_L})$ and
$(\widetilde M^{T_p-S_p}_{M_R}\otimes \widetilde M^{T_p-S_p}_{M_L})$ of the internal structure of an elementary bisemi\-fermion give rise to {\bf the set of equations of its internal dynamics\/}:
\begin{multline*}
(\widetilde M^{T_p-S_p}_{ST_R}\oplus \widetilde M^{T_p-S_p}_{MG_R}
\oplus \widetilde M^{T_p-S_p}_{M_R} ) \otimes
(\widetilde M^{T_p-S_p}_{ST_L}\oplus \widetilde M^{T_p-S_p}_{MG_L}
\oplus \widetilde M^{T_p-S_p}_{M_L} )\\
= (\widetilde M^{T_p-S_p}_{ST_R}\otimes \widetilde M^{T_p-S_p}_{ST_L})
\oplus (\widetilde M^{T_p-S_p}_{MG_R}\otimes \widetilde M^{T_p-S_p}_{MG_L})
\oplus (\widetilde M^{T_p-S_p}_{M_R}\otimes \widetilde M^{T_p-S_p}_{M_L})\\
\qquad \qquad \oplus (\widetilde M^{T_p-S_p}_{ST_R}\otimes \widetilde M^{T_p-S_p}_{MG_L})
\oplus (\widetilde M^{T_p-S_p}_{ST_R}\otimes \widetilde M^{T_p-S_p}_{M_L})
\oplus (\widetilde M^{T_p-S_p}_{MG_R}\otimes \widetilde M^{T_p-S_p}_{ST_L})\\
\qquad \qquad \oplus (\widetilde M^{T_p-S_p}_{MG_R}\otimes \widetilde M^{T_p-S_p}_{M_L})
\oplus (\widetilde M^{T_p-S_p}_{M_R}\otimes \widetilde M^{T_p-S_p}_{ST_L})
\oplus (\widetilde M^{T_p-S_p}_{M_R}\otimes \widetilde M^{T_p-S_p}_{MG_L})\\
=0\end{multline*}
where the set of six mixed bisemisheaves
$(\widetilde M^{T_p-S_p}_{ST_R}\otimes \widetilde M^{T_p-S_p}_{ST_L})
\cdots
 (\widetilde M^{T_p-S_p}_{M_R}\otimes \widetilde M^{T_p-S_p}_{MG_L})$
generates the interactions between the right and left semisheaves of different levels ``~$ST$~'', ``~$MG$~'' and ``~$M$~''.

\item {\bfseries\boldmath If the interactions between the right and left internal semistructures ``~$ST$~'', ``~$MG$~'' and ``~$M$~'' are assumed to be negligible, then the equations of the internal dynamics of a bisemifermion are\/}:
\be\label{eq:*}
(\widetilde M^{T_p-S_p}_{ST_R}\otimes \widetilde M^{T_p-S_p}_{ST_L})
\oplus (\widetilde M^{T_p-S_p}_{MG_R}\otimes \widetilde M^{T_p-S_p}_{MG_L})
= - (\widetilde M^{T_p-S_p}_{M_R}\otimes \widetilde M^{T_p-S_p}_{M_L})
\tag{$*$} \ee
\Ee}
\vskip 11pt

\bpr
\Bi
\item This proposition is a direct consequence of the preceding sections.

\item The interactions between the right and left internal semisheaves
``~$ST$~'', ``~$MG$~'' and ``~$M$~'' are especially responsible for the internal angular momentum of the considered bisemifermion as developed in \cite{Pie5}.

\item If the interactions between the right and left internal semisheaves are assumed to be negligible, then {\bf the equations of the internal dynamics\/}:
\be
(\widetilde M^{T_p-S_p}_{ST_R}\otimes \widetilde M^{T_p-S_p}_{ST_L})
\oplus (\widetilde M^{T_p-S_p}_{MG_R}\otimes \widetilde M^{T_p-S_p}_{MG_L})
= - (\widetilde M^{T_p-S_p}_{M_R}\otimes \widetilde M^{T_p-S_p}_{M_L})
\tag{$*$} \ee
describe the generation of the matter field
$(\widetilde M^{T_p-S_p}_{M_R}\otimes \widetilde M^{T_p-S_p}_{M_L})$ of a bisemifemion from its own vacuum fields given the lefthand side of \eqref{eq:*}.\epr
\Ei
\vskip 11pt

\section{Extended bilinear Hilbert spaces of internal structures}

According to \cite{Pie3}, each
``~$ST$~'', ``~$MG$~'' or ``~$M$~'' field in \eqref{eq:*} is an {\bf operator valued string field\/} on the corresponding string field which defines an extended bilinear Hilbert space.

Indeed, if we apply a $(B_L\circ p_L)$ map on the string fields
$(\widetilde M^{T-S}_{ST_R}\otimes \widetilde M^{T-S}_{ST_L})$~,
$(\widetilde M^{T-S}_{MG_R}\otimes \widetilde M^{T-S}_{MG_L})$ and
$(\widetilde M^{T-S}_{M_R}\otimes \widetilde M^{T-S}_{M_L})$ in such a way that:
\Bi
\item $p_L$ be a projective map projecting each right semifield on its corresponding left equivalent,
\item $B_L$ be a bijective isometric map sending each covariant element into its contravariant equivalent,
\Ei
then they are transformed into:
\begin{alignat*}{3}
(B^{(ST)}_L\circ p^{(ST)}_L) &: 
\quad \widetilde M^{T-S}_{ST_R}\otimes \widetilde M^{T-S}_{ST_L} &\To&
H^+_{ST}= \widetilde M^{T-S}_{ST_{L_R}}\otimes \widetilde M^{T-S}_{ST_L}\\
(B^{(MG)}_L\circ p^{(MG)}_L) &: 
\quad \widetilde M^{T-S}_{MG_R}\otimes \widetilde M^{T-S}_{MG_L} &\To&
H^+_{MG}= \widetilde M^{T-S}_{MG_{L_R}}\otimes \widetilde M^{T-S}_{MG_L}\\
(B^{(M)}_L\circ p^{(M)}_L) &: 
\quad \widetilde M^{T-S}_{M_R}\otimes \widetilde M^{T-S}_{M_L} &\To&
H^+_{M}= \widetilde M^{T-S}_{M_{L_R}}\otimes \widetilde M^{T-S}_{M_L}\end{alignat*}
where $H^+_{ST}$~, $H^+_{MG}$ and $H^+_M$ are the extended bilinear Hilbert spaces of the internal vacuum, middle ground and mass structures of a bisemifermion if complete internal bilinear forms are defined on them.
\vskip 11pt

\section[Actions of bioperators on $H^+_{ST}$~, $H^+_{MG}$ and $H^+_M$]{\bbf Actions of bioperators on $H^+_{ST}$~, $H^+_{MG}$ and $H^+_M$}

On the extended bilinear Hilbert spaces $H^+_{ST}$~, $H^+_{MG}$ and $H^+_M$~, we have the actions of the bioperators:
\begin{alignat*}{3}
T^{T-S}_{R;ST}\otimes T^{T-S}_{L;ST}&: \quad
\widetilde M^{T-S}_{ST_{L_R}}\otimes \widetilde M^{T-S}_{ST_L} &\To&
\widetilde M^{T_p-S_p}_{ST_{L_R}}\otimes \widetilde M^{T_p-S_p}_{ST_L} \\
DT^{T-S}_{R;MG}\otimes DT^{T-S}_{L;MG}&: \quad
\widetilde M^{T-S}_{MG_{L_R}}\otimes \widetilde M^{T-S}_{MG_L} &\To&
\widetilde M^{T_p-S_p}_{MG_{L_R}}\otimes \widetilde M^{T_p-S_p}_{MG_L} \\
DT^{T-S}_{R;M}\otimes DT^{T-S}_{L;M}&: \quad
\widetilde M^{T-S}_{M_{L_R}}\otimes \widetilde M^{T-S}_{M_L} &\To&
\widetilde M^{T_p-S_p}_{M_{L_R}}\otimes \widetilde M^{T_p-S_p}_{M_L} \end{alignat*}
sending the internal vacuum, middle ground and mass bisemisheaves
$\widetilde M^{T-S}_{ST_{L_R}}\otimes \widetilde M^{T-S}_{ST_L}$~,
$\widetilde M^{T-S}_{MG_{L_R}}\otimes \widetilde M^{T-S}_{MG_L}$ and
$\widetilde M^{T-S}_{M_{L_R}}\otimes \widetilde M^{T-S}_{M_L}$ into their perverse equivalents.

The bioperators are explicitly given by:
\begin{multline*}
T^{T-S}_{R;ST}\otimes T^{T-S}_{L;ST}
= \L( -i\ \F{\hbar_{ST}}{c_{t\to r;ST}}\ 
\L\{s_{0_R}\ dt_0;s_{x_R}\ dx, s_{y_R}\ dy,s_{z_R}\ dz\R\}\R)\\
\otimes \L( -i\ \F{\hbar_{ST}}{c_{t\to r;ST}}\ \L\{s_{0_L}\ dt_0;s_{x_L}\ dx, s_{y_L}\ dy,s_{z_L}\ dz\R\}\R)\end{multline*}
\begin{multline*}
DT^{T-S}_{R;MG}\otimes DT^{T-S}_{L;MG}
= \L( -i\ \F{\hbar_{MG}}{c_{t\to r;MG}}\ \L\{s_{0_R}\ \F\partial{\partial t_0};s_{x_R}\ \F\partial{\partial x}, s_{y_R}\ \F\partial{\partial y},s_{z_R}\ \F\partial{\partial z}\R\}\R)\\
\otimes \L( -i\ \F{\hbar_{MG}}{c_{t\to r;MG}}\ \L\{s_{0_L}\ \F\partial{\partial t_0};s_{x_L}\ \F\partial{\partial x}, s_{y_L}\ \F\partial{\partial y},s_{z_L}\ \F\partial{\partial z}\R\}\R)\end{multline*}
\begin{multline*}
DT^{T-S}_{R;M}\otimes DT^{T-S}_{L;M}
= \L( -i\ \F{\hbar_{M}}{c_{t\to r;M}}\ \L\{s_{0_R}\ \F\partial{\partial t_0};s_{x_R}\ \F\partial{\partial x}, s_{y_R}\ \F\partial{\partial y},s_{z_R}\ \F\partial{\partial z}\R\}\R)\\
\otimes \L( -i\ \F{\hbar_{M}}{c_{t\to r;M}}\ \L\{s_{0_L}\ \F\partial{\partial t_0};s_{x_L}\ \F\partial{\partial x}, s_{y_L}\ \F\partial{\partial y},s_{z_L}\ \F\partial{\partial z}\R\}\R)\end{multline*}
where $s_0$ and $(s_x, s_y, s_z)$ are the direction cosines of the unit vectors $\vec s_{t_0}$ and $\vec s_r$ referring to the spin and allowing to defined directional gradients
$\vec s_{t_0}\ \F\partial{\partial t_0}$ and $\vec s_r\ \nabla $~.
\\
(More concretely, the function $\phi (t_0)$ is said to be derivable at $t_0$ in the direction $s$ if $\lim\limits_{\varepsilon \to 0}\ \F{\phi (t_0+\varepsilon s)-\phi (t_0)}\varepsilon $ exists).\vskip 11pt

\section{Proposition}

{\em
\Bi
\item Let $r=q+p$ be the number of algebraic conjugacy (or equivalence) classes of time and space, the indices of time and space varying separately according to $1\le \mu \le q$ and $1\le \nu \le p$ and commonly according to $1\le\sigma \le r$~.

\item Let
\begin{alignat*}{5}
&\bullet &&&\L\{ \widetilde M_{ST_{v_{\sigma ,m_\sigma }}}=
 \widetilde M^T_{ST_{v_{\mu ,m_\mu }}} +  \widetilde M^S_{ST_{v_{\nu ,m_\nu }}}\R\}^r_{\sigma =1} &\subset \widetilde M^{T-S}_{ST_L}\\
 && \quad\Bigg(\text{resp.} \quad
 &&\L\{ \widetilde M_{ST_{\o v_{\sigma ,m_\sigma }}}=
 \widetilde M^T_{ST_{\o v_{\mu ,m_\mu }}} +  \widetilde M^S_{ST_{\o v_{\nu ,m_\nu }}}\R\}^r_{\sigma =1} &\subset \widetilde M^{T-S}_{ST_R}\ \Bigg)\\[11pt]
&\bullet &&&\L\{ \widetilde M_{MG_{v_{\sigma ,m_\sigma }}}=
 \widetilde M^T_{MG_{v_{\mu ,m_\mu }}} +  \widetilde M^S_{MG_{v_{\nu ,m_\nu }}}\R\}^r_{\sigma =1} &\subset \widetilde M^{T-S}_{MG_L}\\
 && \quad\Bigg(\text{resp.} \quad
 && \L\{ \widetilde M_{MG_{\o v_{\sigma ,m_\sigma }}}=
 \widetilde M^T_{MG_{\o v_{\mu ,m_\mu }}} +  \widetilde M^S_{MG_{\o v_{\nu ,m_\nu }}}\R\}^r_{\sigma =1} &\subset \widetilde M^{T-S}_{MG_R}\ \Bigg)\\[11pt]
 \noalign{\newpage}
&\bullet &&&\L\{ \widetilde M_{M_{v_{\sigma ,m_\sigma }}}=
 \widetilde M^T_{M_{v_{\mu ,m_\mu }}} +  \widetilde M^S_{M_{v_{\nu ,m_\nu }}}\R\}^r_{\sigma =1} &\subset \widetilde M^{T-S}_{M_L}\\
 && \quad\Bigg(\text{resp.} \quad
 && \L\{ \widetilde M_{M_{\o v_{\sigma ,m_\sigma }}}=
 \widetilde M^T_{M_{\o v_{\mu ,m_\mu }}} +  \widetilde M^S_{M_{\o v_{\nu ,m_\nu }}}\R\}^r_{\sigma =1} &\subset \widetilde M^{T-S}_{M_R}\ \Bigg)\end{alignat*}
 be the set of sections of the space-time \lr semisheaves of the internal vacuum (``~$ST$~''),
 middle ground (``~$MG$~'') and mas (``~$M$~'') structures of a \lr semifermion.
 
 \item Let
 \begin{alignat*}{5}
&\bullet&&& T_{ST_L} &= -i\ \F{\hbar_{ST}}{c_{t\to r;ST}}\ \L(\vec s_{t_0}\ dt_0 + \vec s_{r_x}\ dx + \vec s_{r_y}\ dy + \vec s_{r_z}\ dz\R)\\
&&\quad\Bigg(\text{resp.} \quad 
&&T_{ST_R} &= +i\ \F{\hbar_{ST}}{c_{t\to r;ST}}\ \L(\vec s_{t_0}\ dt_0 + \vec s_{r_x}\ dx + \vec s_{r_y}\ dy + \vec s_{r_z}\ dz\R)\Bigg)\\
&\bullet& &&DT_{MG_L} &= -i\ \F{\hbar_{MG}}{c_{t\to r;MG}}\ \L(\vec s_{t_0}\ \F\partial{\partial t_0} + \vec s_{r_x}\ \F\partial{\partial x}+ \vec s_{r_y}\ \F\partial{\partial y} + \vec s_{r_z}\ \F\partial{\partial z}\R)\\
&&\quad\Bigg(\text{resp.} \quad 
&&DT_{MG_R} &= +i\ \F{\hbar_{MG}}{c_{t\to r;MG}}\ \L(\vec s_{t_0}\ \F\partial{\partial t_0} + \vec s_{r_x}\ \F\partial{\partial x}+ \vec s_{r_y}\ \F\partial{\partial y} + \vec s_{r_z}\ \F\partial{\partial z}\R)\Bigg)\\
&\bullet& &&DT_{M_L} &= -i\ \F{\hbar_{M}}{c_{t\to r;M}}\ \L(\vec s_{t_0}\ \F\partial{\partial t_0} + \vec s_{r_x}\ \F\partial{\partial x}+ \vec s_{r_y}\ \F\partial{\partial y} + \vec s_{r_z}\ \F\partial{\partial z}\R)\\
&&\quad\Bigg(\text{resp.} \quad 
&&DT_{M_R} &= +i\ \F{\hbar_{M}}{c_{t\to r;M}}\ \L(\vec s_{t_0}\ \F\partial{\partial t_0} + \vec s_{r_x}\ \F\partial{\partial x}+ \vec s_{r_y}\ \F\partial{\partial y} + \vec s_{r_z}\ \F\partial{\partial z}\R)\Bigg)
\end{alignat*}
be the corresponding differential operators acting on these sets of sections.
\Ei

Then, {\bfseries\boldmath the equations \eqref{eq:*} of the internal dynamics of a bisemifermion can be put in the equivalent form\/}:
\begin{multline*}
\L[ T_{ST_R}(\widetilde M_{ST_{\o v_{\sigma ,m_\sigma }}}) \otimes 
T_{ST_L}(\widetilde M_{ST_{v_{\sigma ,m_\sigma }}})\R]
+ \L[ DT_{MG_R}(\widetilde M_{MG_{\o v_{\sigma ,m_\sigma }}}) \otimes 
DT_{MG_L}(\widetilde M_{MG_{v_{\sigma ,m_\sigma }}})\R]\\
= -\L[ DT_{M_R}(\widetilde M_{M_{\o v_{\sigma ,m_\sigma }}}) \otimes 
DT_{M_L}(\widetilde M_{M_{v_{\sigma ,m_\sigma }}})\R]\;, \quad \forall\ \sigma ,m_\sigma \ , \; 1\le \sigma \le r\;.\end{multline*}
}\vskip 11pt

\bpr Indeed, it appears from the preceding developments that the actions of the bioperators 
$(T_{ST_R} \otimes 
T_{ST_L})$~, $(DT_{MG_R} \otimes 
DT_{MG_L})$ and $(DT_{M_R} \otimes 
DT_{M_L})$ on the corresponding bisections, as developed above, transform these into bisections of the corresponding perverse bisemisheaves and give rise to the equations \eqref{eq:*} of the internal dynamics of a bisemifermion.\epr
\vskip 11pt

\section{Corollary}

{\em Let $T_{ST_{L_R}}$~,   $DT_{MG_{L_R}}$ and $DT_{M_{L_R}}$ be the adjoint operators of
$T_{ST_R}$~,   $DT_{MG_R}$ and $DT_{M_R}$ respectively:
\[ T_{ST_{L_R}}=T\tt _{ST_R}\;, \quad 
DT_{MG_{L_R}}=DT\tt _{MG_R}\quad \and \quad DT_{M_{L_R}}=DT\tt _{M_R}\;.\]

Let $\widetilde M_{\o{ST}_{\o v_{\sigma ,m_\sigma }}}$~, 
$\widetilde M_{\o{MG}_{\o v_{\sigma ,m_\sigma }}}$ and
$\widetilde M_{\o{M}_{\o v_{\sigma ,m_\sigma }}}$ be the sections of the semisheaves
$\widetilde M^{T-S}_{ST_R}$~,  $\widetilde M^{T-S}_{MG_R}$ and $\widetilde M^{T-S}_{M_R}$ projected onto the sections of the corresponding left semisheaves according to:
\begin{alignat*}{3}
B^{(ST)}_L\circ p_L^{(ST)} &: \quad \widetilde M_{{ST}_{\o v_{\sigma ,m_\sigma }}}
&\To & \widetilde M_{\o{ST}_{\o v_{\sigma ,m_\sigma }}}\;, \\
B^{(MG)}_L\circ p_L^{(MG)} &: \quad \widetilde M_{{MG}_{\o v_{\sigma ,m_\sigma }}}
&\To & \widetilde M_{\o{MG}_{\o v_{\sigma ,m_\sigma }}}\;, \\
B^{(M)}_L\circ p_L^{(M)} &: \quad \widetilde M_{{M}_{\o v_{\sigma ,m_\sigma }}}
&\To & \widetilde M_{\o{M}_{\o v_{\sigma ,m_\sigma }}}\;.\end{alignat*}

Then, the equations on  the sections of the internal dynamics of a bisemifermion are transformed under the isomorphism $ID_{R\to L_R}$ into:
\begin{multline*}
ID_{R\to L_R} : \quad [T_{ST_R}( \widetilde M_{{ST}_{\o v_{\sigma ,m_\sigma }}}
) \otimes T_{ST_L}( \widetilde M_{{ST}_{v_{\sigma ,m_\sigma }}}
)]\\
\begin{aligned}
&+ [DT_{MG_R}( \widetilde M_{{MG}_{\o v_{\sigma ,m_\sigma }}}
) \otimes DT_{MG_L}( \widetilde M_{{MG}_{v_{\sigma ,m_\sigma }}}
)]\\
& \quad = - [DT_{M_R}( \widetilde M_{{M}_{\o v_{\sigma ,m_\sigma }}}
) \otimes DT_{M_L}( \widetilde M_{{M}_{v_{\sigma ,m_\sigma }}}
)]\\
& \qquad \To     [(T_{ST_{L_R}} \times  T_{ST_L}) ( \widetilde M_{\o{ST}_{\o v_{\sigma ,m_\sigma }}}
 \times \widetilde M_{{ST}_{v_{\sigma ,m_\sigma }}}
)]\\
& \qquad \qquad + [(DT_{MG_{L_R}} \times  DT_{MG_L}) ( \widetilde M_{\o{MG}_{\o v_{\sigma ,m_\sigma }}}
 \times \widetilde M_{{MG}_{v_{\sigma ,m_\sigma }}}
)]
\end{aligned}\\
= -[(DT_{M_{L_R}} \times  DT_{M_L}) ( \widetilde M_{\o{M}_{\o v_{\sigma ,m_\sigma }}}
 \times \widetilde M_{{M}_{v_{\sigma ,m_\sigma }}}
)]\;.\end{multline*}
}\vskip 11pt

\bpr This isomorphism $ID_{R\to L_R}$ transforms the (tensor) product of the right action of
$DT_{M_R}$ on $\widetilde M_{M_{\o v_{\sigma ,m_\sigma }}}$ by the left action of
$DT_{M_L}$ on $\widetilde M_{M_{v_{\sigma ,m_\sigma }}}$ into the biaction of
$(DT_{M_{L_R}}\times DT_{M_L})$ onto the product of the sections
$(\widetilde M_{\o M_{\o v_{\sigma ,m_\sigma }}} \times \widetilde M_{M_{v_{\sigma ,m_\sigma }}})$~~, and so on for the two other levels ``~$ST$~'' and ``~$MG$~''.

And, the products of the following sections belong to the extended bilinear Hilbert spaces:
\[
\widetilde M_{\o{ST}_{\o v_{\sigma ,m_\sigma }}} \times \widetilde M_{{ST}_{v_{\sigma ,m_\sigma }}} \in H^+_{ST}\;, \quad
\widetilde M_{\o{MG}_{\o v_{\sigma ,m_\sigma }}} \times \widetilde M_{{MG}_{v_{\sigma ,m_\sigma }}} \in H^+_{MG}\;, \quad
\widetilde M_{\o{M}_{\o v_{\sigma ,m_\sigma }}} \times \widetilde M_{{M}_{v_{\sigma ,m_\sigma }}} \in H^+_{M}\;, \]
according to section 3.9.\epr

%% file: CPTrack4.tex
\chapter{Equivalence between the equations of general relativity and the equations of the internal dynamics of bisemiparticles}

\thispagestyle{empty}

The aim of this chapter consists in generalizing the equations of the internal dynamics to a set of $J$ (bisemi)particles, and, more particularly, to a set of $J$ elementary (bisemi)fermions and in pointing out that they are in one-to-one correspondence with the equations of general relativity.
\vskip 11pt

\section{The internal structure of a set of bisemifermions}

\Bi
\item It was seen in chapter 3 that the time or space string field of the internal vacuum, middle ground or mass structure of a bisemifermion is given by the representation space (in bisemisheaf) of the bilinear algebraic semigroup $\GL_2(L_{\o v}\times L_v)$~.
\vskip 11pt

\item Generalizing to a set of $J$ bisemifermions, we have to take into account the partition $2J= 2_1+\cdots+2_i+\cdots +2_J$ of $2J$ in such a way that the ``~$ST$~'', ``~$MG$~'' or ``~$M$~'' string field of time or space of them be given by {\bbf the completely reducible non orthogonal representation space $\Repsp(\GL_2(L_{\o v}\times L_v))$\/} of the bilinear algebraic semigroup $\GL_{2J}(L_{\o v}\times L_v)$ of dimension $2J$~, as introduced in \cite{Pie2}.

This representation space $\Repsp(\GL_{2J}(L_{\o v}\times L_v))$ {\bf decomposes non orthogonally according to\/}:
\[
\Repsp(\GL_{2J}(L_{\o v}\times L_v)) = \bplus^J_{i=1} 
\Repsp(\GL_{2_i}(L_{\o v}\times L_v))
\bplus^J_{i\neq j=1} 
\Repsp(T^t_{2_i}(L_{\o v})\times T_{2_j}(L_v))\]
where the {\bf off-diagonal representation spaces $ 
\Repsp(T^t_{2_i}(L_{\o v})\times T_{2_j}(L_v))$ are responsible for the generation of gravito-electro-magnetic fields of interaction\/} between bisemifermions as developed in \cite{Pie5}.
\vskip 11pt

\item Referring to Proposition 3.7, we see that the internal vacuum  (``~$ST$~'') field structure of space-time of $J$ interacting bisemifermions is given by the bisemisheaves:
\begin{align*}
(\widetilde M^{T-S}_{ST_{R_J}}\otimes \widetilde M^{T-S}_{ST_{L_J}})
&= \Repsp\theta (\GL_{2J(T+S)}(L_{\o v}\times L_v))\\
&= \bplus^J_{i=1}
(\widetilde M^{T-S}_{ST_{R_i}}\otimes \widetilde M^{T-S}_{ST_{L_i}})
\bplus^J_{i\neq j=1}
(\widetilde M^{T-S}_{ST_{R_i}}\otimes \widetilde M^{T-S}_{ST_{L_j}})\;.\end{align*}
\vskip 11pt

\item By versal deformations and blowups of these, the ``internal vacuum'' string fields
$
(\widetilde M^{T-S}_{ST_{R_J}}\otimes \widetilde M^{T-S}_{ST_{L_J}})$ of a set of $J$ interacting bisemifermions generate {\bf the two covering ``middle ground'' and ``mass'' string fields of space-time\/}:
\begin{align*}
(\widetilde M^{T-S}_{MG_{R_J}}\otimes \widetilde M^{T-S}_{MG_{L_J}})
&= \bplus^J_{i=1}
(\widetilde M^{T-S}_{MG_{R_i}}\otimes \widetilde M^{T-S}_{MG_{L_i}})
\bplus^J_{i\neq j=1}
(\widetilde M^{T-S}_{MG_{R_i}}\otimes \widetilde M^{T-S}_{MG_{L_j}}) \\
\noalign{\qquad \qquad and}
(\widetilde M^{T-S}_{M_{R_J}}\otimes \widetilde M^{T-S}_{M_{L_J}})
&= \bplus^J_{i=1}
(\widetilde M^{T-S}_{M_{R_i}}\otimes \widetilde M^{T-S}_{M_{L_i}})
\bplus^J_{i\neq j=1}
(\widetilde M^{T-S}_{M_{R_i}}\otimes \widetilde M^{T-S}_{M_{L_j}}) \end{align*}
in such a way that we have the embeddings:
\[ (\widetilde M^{T-S}_{ST_{R_J}}\otimes \widetilde M^{T-S}_{ST_{L_J}})
\subset (\widetilde M^{T-S}_{MG_{R_J}}\otimes \widetilde M^{T-S}_{MG_{L_J}})
\subset
(\widetilde M^{T-S}_{M_{R_J}}\otimes \widetilde M^{T-S}_{M_{L_J}})\;.\]
\Ei
\vskip 11pt

\section{Proposition}

{\em Let $(\widetilde M^{T_p-S_p}_{ST_{R_J}}\otimes \widetilde M^{T_p-S_p}_{ST_{L_J}})
$~, $(\widetilde M^{T_p-S_p}_{MG_{R_J}}\otimes \widetilde M^{T_p-S_p}_{MG_{L_J}})
$ and $
(\widetilde M^{T_p-S_p}_{M_{R_J}}\otimes \widetilde M^{T-S}_{M_{L_J}})$ be the perverse bisemisheaves of the internal structures  ``~$ST$~'', ``~$MG$~'' and ``~$M$~'' of a set of $J$ interacting bisemifermions.

If the interactions between the right and left internal semifields
``~$ST$~'', ``~$MG$~'' and ``~$M$~'' are negligible, then {\bfseries\boldmath the equations of the internal dynnamics of this set of $J$ bisermifermions are\/}:
\[ (\widetilde M^{T_p-S_p}_{ST_{R_J}}\otimes \widetilde M^{T_p-S_p}_{ST_{L_J}})
\oplus (\widetilde M^{T_p-S_p}_{MG_{R_J}}\otimes \widetilde M^{T_p-S_p}_{MG_{L_J}})
=-
(\widetilde M^{T_p-S_p}_{M_{R_J}}\otimes \widetilde M^{T-S}_{M_{L_J}})\;.\]
}
\vskip 11pt

\bpr
\Bi
\item The perverse bisemisheaves
$(\widetilde M^{T_p-S_p}_{ST_{R_J}}\otimes \widetilde M^{T_p-S_p}_{ST_{L_J}})
$~, $(\widetilde M^{T_p-S_p}_{MG_{R_J}}\otimes \widetilde M^{T_p-S_p}_{MG_{L_J}})
$ and $
(\widetilde M^{T_p-S_p}_{M_{R_J}}\otimes \widetilde M^{T-S}_{M_{L_J}})$ are obtained from the corresponding bisemisheaves $
(\widetilde M^{T-S}_{ST_{R_J}}\otimes \widetilde M^{T-S}_{ST_{L_J}})
$~,\linebreak $(\widetilde M^{T-S}_{MG_{R_J}}\otimes \widetilde M^{T-S}_{MG_{L_J}})
$ and $
(\widetilde M^{T-S}_{M_{R_J}}\otimes \widetilde M^{T-S}_{M_{L_J}})$ by the respective actions of the biopearators $(T_{ST_L}\otimes T_{ST_R})$~, 
$(DT_{MG_L}\otimes DT_{MG_R})$ and $(DT_{M_R}\otimes DT_{M_L})$ introduced in section 3.11.

\item The equations of the internal dynamics of a set of $J$ bisemifermions are a generalization of the equations \eqref{eq:*} of proposiiton 3.8.\epr
\Ei
\vskip 11pt

\section{Proposition}

{\em
\Bi
\item Let $r_i=q_i+p_i$~, $1\le i\le J$~, be the numbers of algebraic conjugacy classes of time and space varying commonly according to $1\le \sigma _i\le r_i$~.

\item Let \bt[t]{llrl}
&$\{\widetilde M_{ST_{v_{\sigma_i,m_{\sigma _i}}}}\}^{r_i}_{\sigma _i=1} \subset \widetilde M^{T-S}_{ST_{L_i}}$ & \qquad (resp. &
\quad $\{\widetilde M_{ST_{\o v_{\sigma_i,m_{\sigma _i}}}}\}^{r_i}_{\sigma _i=1} \subset \widetilde M^{T-S}_{ST_{R_i}}$~)\\
&$\{\widetilde M_{MG_{v_{\sigma_i,m_{\sigma _i}}}}\}^{r_i}_{\sigma _i=1} \subset \widetilde M^{T-S}_{MG_{L_i}}$ & \qquad (resp. &
\quad $\{\widetilde M_{MG_{\o v_{\sigma_i,m_{\sigma _i}}}}\}^{r_i}_{\sigma _i=1} \subset \widetilde M^{T-S}_{MG_{R_i}}$~)\\
and &$\{\widetilde M_{M_{v_{\sigma_i,m_{\sigma _i}}}}\}^{r_i}_{\sigma _i=1} \subset \widetilde M^{T-S}_{M_{L_i}}$ & \qquad (resp. &
\quad $\{\widetilde M_{M_{\o v_{\sigma_i,m_{\sigma _i}}}}\}^{r_i}_{\sigma _i=1} \subset \widetilde M^{T-S}_{M_{R_i}}$~)\te

be the set of sections of the space-time \lr semisheaves of the ``internal vacuum'' (~$ST$~), ``middle ground'' (~$MG$~) and ``mass'' (~$M$~) structures of the $J$~, $1\le i\le J$~, \lr considered semifermions.
\Ei

Then, {\bfseries the equations 
\[ (\widetilde M^{T_p-S_p}_{ST_{R_J}}\otimes \widetilde M^{T_p-S_p}_{ST_{L_J}})
\oplus (\widetilde M^{T_p-S_p}_{MG_{R_J}}\otimes \widetilde M^{T-S}_{MG_{L_J}})
=-
(\widetilde M^{T_p-S_p}_{M_{R_J}}\otimes \widetilde M^{T-S}_{M_{L_J}})\]
of the internal dynamics of a set of {\boldmath $J$} bisemifermions are in one-to-one correspondence with the equations
\begin{multline*}
 [T_{ST_{R_i}}(\widetilde M_{ST_{\o v_{\sigma _i,m_{\sigma _i}}}})
\otimes
T_{ST_{L_j}}(\widetilde M_{ST_{v_{\sigma _j,m_{\sigma _j}}}})]\\
\begin{aligned}
+ &[DT_{MG_{R_i}}(\widetilde M_{MG_{\o v_{\sigma _i,m_{\sigma _i}}}})
\otimes
DT_{ST_{L_j}}(\widetilde M_{MG_{v_{\sigma _j,m_{\sigma _j}}}})]\\
&\qquad =-[DT_{M_{R_i}}(\widetilde M_{M_{\o v_{\sigma _i,m_{\sigma _i}}}})
\otimes
DT_{ST_{L_j}}(\widetilde M_{M_{v_{\sigma _j,m_{\sigma _j}}}})]\end{aligned}\\
\quad \forall\ \sigma _i,\sigma _j\ , \; 1\le \sigma _i,\sigma _j\le r_i,r_j \;{and} \;\forall\ i,j\ \; 1\le i,j\le J\le\infty \end{multline*}
referring to the sections of the considered bisemisheaves\/}.}
\vskip 11pt

\bpr
The one-to-one correspondence between the two types of equations results from:
\Bi
\item[-] a generalization of proposition 3.11 to a set of $J$ bisemifermions.
\item[-] a homomorphism
\begin{multline*}
H_{ST} : \quad
(\widetilde M^{T_p-S_p}_{ST_{R_J}}\otimes \widetilde M^{T_p-S_p}_{ST_{L_J}})
=\bplus^J_{i=1}
(\widetilde M^{T_p-S_p}_{ST_{R_i}}\otimes \widetilde M^{T-S}_{ST_{L_i}})
\bplus^J_{i\neq j=1}
(\widetilde M^{T_p-S_p}_{ST_{R_i}}\otimes \widetilde M^{T-S}_{ST_{L_j}})\\
\To 
[T_{ST_{R_i}}(\widetilde M_{ST_{\o v_{\sigma _i,m_{\sigma _i}}}})
\otimes
T_{ST_{L_j}}(\widetilde M_{ST_{v_{\sigma _j,m_{\sigma _j}}}})]
\end{multline*}
sending the sums $\bplus^J_{i=1}$ and $\bplus^J_{i\neq j=1}$ of bisemisheaves, referring respectively to diagonal and off diagonal interactions between right and left ``~$ST$~'' semistructures of the $J$ bisemifermions, into the set 
$[T_{ST_{R_i}}(\widetilde M_{ST_{\o v_{\sigma _i,m_{\sigma _i}}}})
\otimes
T_{ST_{L_j}}(\widetilde M_{ST_{v_{\sigma _j,m_{\sigma _j}}}})]$ of the corresponding bisections of these bisemishaves. 

\item[-] two similar homomorphisms $H_{MG}$ and $H_M$ referring to the ``~$MG$~'' and ``~$M$~''
structures of the $J$ bisemifermions.\epr
\Ei
\vskip 11pt

\section{Proposition}

{\em 
\Bi
\item Let \bt[t]{llll}
& $B_L^{(ST)}\circ p_L^{(ST)}:$ & \quad 
$\widetilde M_{ST_{\o v_{\sigma _i,m_{\sigma _i}}}}$ & $\To
\widetilde M_{\o{ST}_{\o v_{\sigma _i,m_{\sigma _i}}}}$\\
& $B_L^{(MG)}\circ p_L^{(MG)}:$ & \quad 
$\widetilde M_{MG_{\o v_{\sigma _i,m_{\sigma _i}}}}$ & $\To
\widetilde M_{\o{MG}_{\o v_{\sigma _i,m_{\sigma _i}}}}$\\
and & $B_L^{(M)}\circ p_L^{(M)}:$ & \quad 
$\widetilde M_{M_{\o v_{\sigma _i,m_{\sigma _i}}}}$ & $\To
\widetilde M_{\o{M}_{\o v_{\sigma _i,m_{\sigma _i}}}}$\te

be the maps projecting the sections 
$\widetilde M_{ST_{\o v_{\sigma _i,m_{\sigma _i}}}}$~,
$\widetilde M_{MG_{\o v_{\sigma _i,m_{\sigma _i}}}}$ and
$\widetilde M_{M_{\o v_{\sigma _i,m_{\sigma _i}}}}$ onto their left equivalents.

\item Let $T_{ST_{L_{R_i}}}$~, $DT_{MG_{L_{R_i}}}$ and $DT_{M_{L_{R_i}}}$ be the adjoint operators of $T_{ST_{R_i}}$~, $DT_{MG_{R_i}}$ and $DT_{M_{R_i}}$ respectively:
\[ T_{ST_{L_{R_i}}} = T\tt_{ST_{R_i}} \;, \quad
T_{MG_{L_{R_i}}} = T\tt_{MG_{R_i}} \quad \and \quad T_{M_{L_{R_i}}} = T\tt_{M_{R_i}}\;.\]
\Ei

Then, {\bfseries the equations on the sections of the internal dynamics of a set of  $J$ bisemifermions are transformed under the isomorphisms\/} {\boldmath $ID_{R_J\to L_{R_J}}$} into:
\begin{multline*}
ID_{R_J\to L_{R_J}} : \quad 
[T_{ST_{R_i}} (\widetilde M_{ST_{\o v_{\sigma _i,m_{\sigma _i}}}})
\otimes
T_{ST_{L_j}} (\widetilde M_{ST_{v_{\sigma _j,m_{\sigma _j}}}})]\\
\begin{aligned}
+ & [DT_{MG_{R_i}} (\widetilde M_{MG_{\o v_{\sigma _i,m_{\sigma _i}}}})
\otimes
DT_{MG_{L_j}} (\widetilde M_{MG_{v_{\sigma _j,m_{\sigma _j}}}})]\\
& \quad =-[DT_{M_{R_i}} (\widetilde M_{M_{\o v_{\sigma _i,m_{\sigma _i}}}})
\otimes
DT_{M_{L_j}} (\widetilde M_{M_{v_{\sigma _j,m_{\sigma _j}}}})]
\end{aligned}\\
\To 
[(T_{ST_{L_{R_i}}}\times T_{ST_{L_j}})(\widetilde M_{\o{ST}_{\o v_{\sigma _i,m_{\sigma _i}}}}
\times \widetilde M_{{ST}_{v_{\sigma _i,m_{\sigma _i}}}})]\\
\begin{aligned}
+ & [(DT_{MG_{L_{R_i}}}\times DT_{MG_{L_j}})(\widetilde M_{\o{MG}_{\o v_{\sigma _i,m_{\sigma _i}}}}\times 
\widetilde M_{{MG}_{v_{\sigma _j,m_{\sigma _j}}}})]\\
& \quad =-[(DT_{M_{L_{R_i}}}\times DT_{M_{L_j}})(\widetilde M_{\o{M}_{\o v_{\sigma _i,m_{\sigma _i}}}}\times 
\widetilde M_{{M}_{v_{\sigma _j,m_{\sigma _j}}}})]\;.
\end{aligned}\end{multline*}
}
\vskip 11pt

\bpr This proposition is a generalization of corollary 3.12 to a set of $J$ bisemi-\linebreak fermions.\epr
\vskip 11pt

\section{The small value of the cosmological constant}

The equations obtained under the isomorphism $ID_{R_J\to L_{R_J}}$ and describing the internal dynamics of a set of (bisemi)fermions are rather close to the equations of general relativity as it will be seen in the following sections.

But, one of the big problems of the equations of general relativity in connection with the phenomenology of quantum field theories consists in the small value given to the cosmological constant.  Indeed, in order to avoid a static solution to his equations:
\[ G_{\mu \nu } = 8\pi GT_{\mu \nu }\]
where:
\Bean
\item $G_{\mu \nu }=R_{\mu \nu }-\half\ g_{\mu \nu }R$ \quad with:
\Bi
\item $R_{\mu \nu }$ the Ricci tensor which is the contracted form of the Riemann-Christoffel curvature tensor $R^\lambda _{\mu \lambda \nu }$ by $R_{\mu \nu }= R^\lambda _{\mu \lambda \nu }$~,
\item $R=g^{\mu \nu }R_{\mu \nu }$ the curvature scalar,
\item $g_{\mu \nu }$ the metric tensor of space-time,
\Ei

\item $T_{\mu \nu }$ is the symmetric energy-momentum tensor and $G$ is the gravitational constant,
\Ee

Einstein introduced in these a new term $\lambda g_{\mu \nu }$~, where {\bbf $\lambda $ is the cosmological constant\/}, leading to \cite{Ein2}, \cite{Ein3}:
{\boldmath
\[ \lambda g_{\mu \nu }+G_{\mu \nu }=8\pi GT_{\mu \nu }\;.\]}

The equations $G_{\mu \nu }=8\pi GT_{\mu \nu }$ describe how matter, given by $T_{\mu \nu }$~, generates gravitational forces, characterized by the tensor $G_{\mu \nu }$~, by means of the curvature of the space-time; indeed, the gravitational field is assumed to be represented by the metric tensor itself.

In this context, Zel'dovich \cite{Zel1}, \cite{Zel2} envisaged {\bbf the possible connection between the vacuum energy density of quantum field theories and the Einstein's cosmological constant $\lambda $\/} in such a way that the author proposed in \cite{Pie1} to describe the vacuum energy density $\rho _{\rm vac}$ by
\[ \rho _{\rm vac}\simeq \rho ^{(ST)}_{\rm vac} + \rho^{(MG)} _{\rm vac}\simeq \lambda _{\rm eff}\big/ 8\pi G\]
where:
\Bi
\item the internal vacuum energy density $\rho^{(ST)} _{\rm vac}$ could correspond to $\lambda  /8\pi G$~,
\item $\rho^{(MG)} _{\rm vac}$ would correspond to the middle ground energy density ``~$MG$~''.
\Ei

The revised equations of general relativity can then take the form:
\[ G_{\mu \nu } = 8\pi G\ \L(T_{\mu \nu }-\txt\F\lambda {8\pi G}\ g_{\mu \nu} \R)\]
in such a way that $\lambda g_{\mu \nu} $ could correspond to the internal vacuum energy density $\rho^{(ST)} _{\rm vac}$ which would behave like an ideal fluid with negative pressure $\rho^{(ST)} _{\rm vac}=-p^{(ST)}$ \cite{Pee}, \cite{Lem}.

Considerations on the expansion of the universe allowed Weinberg \cite{Wei1} to show that the effective cosmological constant $\lambda _{\rm eff}$ can be related to the Hubble constant $H_0$ by:
\[ |\lambda _{\rm eff}|\le H_0^2\]
and, also, that:
\[ |\rho _{\rm vac}|\le 10^{-29}{\rm g/cm}^3\simeq 10^{-47}\rm{Gev}\;.\]

So, the present expansion rate of the universe viewed throughout the curvature of space-time by means of the Friedmann's model \cite{Frie} leads to envisage very small values for the vacuum energy density and the related cosmological constant \cite{Wei1}.

This expectation small value of the cosmological constant $\lambda $ must be understood in the frame of general relativity describing gravity by the curvature of space-time by noticing that {\bbf the square root of $\lambda ^{-1}$ is a ``distance'' referring to the domain where the vacuum energy density $\rho^{(ST)} _{\rm vac}$ alters the geometry of space-time by its gravitational effects\/} \cite{Abb} in such a way that the curvature be on an average null.
\vskip 11pt

\section{New interpretation of the general relativity equations}

Thus, the only way to go beyond this problem of the small value of $\lambda $ is to take into account a new interpretation of the equations of general relativity \cite{Ein1}
\[ \lambda g_{\mu \nu }+G_{\mu \nu }=8\pi GT_{\mu \nu }\]
as developed in \cite{Pie5} and \cite{Pie3}.

{\bf This new context is especially characterized by\/}:
\Bean
\item the fact that {\bf the gravitational potential is no more\/} assumed to be {\bbf described by the metric tensor $g_{\mu \nu }$\/} and the gravity is thus not explicitly described by the tensor $G_{\mu \nu }$ when the context of curved space-time geometry remains.

\item {\bbf the composition of ``matter'' given by the three embedded structures $ST\subset MG\subset M$\/} at the elementary particle level, as developed from the beginning of chapter 3, in such a way that:
\Bi
\item the terms $\lambda g_{\mu \nu }+G_{\mu \nu }$ refer to the vacuum structure of matter as it clearly appears from section 4.5 that, when the energy stress tensor of matter $T _{\mu \nu }$ (corresponding to the level ``~$M$~'') is null, then the revised equations of $GR$ in the vacuum are:
\[ \lambda g_{\mu \nu }+G_{\mu \nu }=0\;.\]
\item\Bi
\item the term $\lambda g_{\mu \nu }$ would be associated to the ``~$ST$~'' internal vacuum substructures;
\item the term $G_{\mu \nu }$ would correspond to the ``~$MG$~'' middle ground vacuum substructures;
\item the term $8\pi GT_{\mu \nu }$ would correspond to the ``~$M$~'' mass boundary structures.
\Ei\Ei\Ee

It then appears that the vacuum, considered at the elementary particle level, must be composed of:
\Be
\item the ``internal vacuum'' substructures ``~$ST$~'' of which nature is of space-time type having a dynamical and expansive aspect.
\item the ``middle ground'' substructures generated from the corresponding internal ``~$ST$~'' substructures and having a contracting aspect allowing to confine ``~$ST$~'' substructures inside the elementary particles.
\Ee

{\bf This elementary particle vacuum then corresponds to the vacuum of the quantum field theories allowing to generate the particle masses\/} \cite{Par} (i.e. the ``~$M$~'' boundary structures, in the context of AQT, from the degenerate singularities on the ``middle ground'' substructures) and would be associated with the map:
\[ G_M: \quad \lambda g_{\mu \nu }+G_{\mu \nu }=0\To \lambda g_{\mu \nu }+G_{\mu \nu }=8\pi GT_{\mu \nu }\]
in the frame of general relativity equations.
\vskip 11pt

This new interpretation of general relativity, connecting its vacuum energy with that of quantum field theories, was foreboded by Sakharov \cite{Sak} who claimed that gravity is not a fundamental quantum field but an induced quantum effect caused by an interaction of quantum vacuum fluctuations with space-time curvature.

In other terms, the inhomogeneity of vacuum fluctuations induces the Riemann space-time geometry \cite{Gli}, \cite{Ban} which does not describe ontologically gravity.
\vskip 11pt

\section[Compactification of the ``~$ST$~'', ``~$MG$~'' and ``~$M$~'' structures]{\bbf Compactification of the ``~$ST$~'', ``~$MG$~'' and ``~$M$~'' structures}

It remains to prove that a one-to-one correspondence, given by the isomorphism $I_{QT\to GR}$~, exists between the equations of general relativity and the string field equations on the sections of the internal dynamics of a set of (bisemi)fermions (or, more generally, of a set of (bisemi)particles) as given in proposition 4.4:
\begin{multline*}
I_{QT\to GR} : \quad 
[(T_{ST_{L_{R_i}}}\times T_{ST_{L_j}})(\widetilde M_{\o{ST}_{\o v_{\sigma _i,m_{\sigma _i}}}}
\times \widetilde M_{{ST}_{\o v_{\sigma _j,m_{\sigma _j}}}})]\\
\begin{aligned}
+ & [(DT_{MG_{L_{R_i}}}\times DT_{MG_{L_j}})(\widetilde M_{\o{MG}_{\o v_{\sigma _i,m_{\sigma _i}}}}\times 
\widetilde M_{{MG}_{\o v_{\sigma _j,m_{\sigma _j}}}})]\\
& \quad =-[(DT_{M_{L_{R_i}}}\times DT_{M_{L_j}})(\widetilde M_{\o{M}_{\o v_{\sigma _i,m_{\sigma _i}}}}\times 
\widetilde M_{{M}_{\o v_{\sigma _j,m_{\sigma _j}}}})]
\end{aligned}\\
\To \lambda g_{\mu \nu }+G_{\mu \nu }=8\pi GT_{\mu \nu }\;.\end{multline*}
To reach this objective, it is necessary to:
\Be
\item compactify the diagonal perverse semisheaves
$(T_{ST_{L_{R_i}}} \widetilde M_{\o{ST}_{\o v_{\sigma _i,m_{\sigma _i}}}})$
and $(T_{ST_{L_i}} \widetilde M_{{ST}_{\o v_{\sigma _i,m_{\sigma _i}}}})$ of the internal vacua
``~$ST$~'' of the ``~$J$~'' considered bisemifermions as it was done in section 2.10 in order to obtain four-dimensional (1 dimension of time and 3 dimensions of space) perverse right (resp. left) compactified semisheaves according to the compactification map:
\begin{alignat*}{3}
c^{t-r}_{ST_R} &: \quad
 T_{ST_{L_{R_i}}} \widetilde M_{\o{ST}_{\o v_{\sigma _i,m_{\sigma _i}}}}
 & \To &
 T_{ST_{L_{R_i}}} \widetilde M^c_{\o{ST}_{\o v_{\sigma _i,m_{\sigma _i}}}}\\
 \text{(resp.} \quad 
 c^{t-r}_{ST_L} &: \quad
 T_{ST_{L_i}} \widetilde M_{{ST}_{v_{\sigma _i,m_{\sigma _i}}}}
 & \To &
 T_{ST_{L_i}} \widetilde M^c_{{ST}_{v_{\sigma _i,m_{\sigma _i}}}}\ )\;.\end{alignat*}
 
 \item compactify similarly the diagonal perverse semisheaves of the right and left ``middle ground'' (~$MG$~) and ``mass'' (~$M$~) structures of the considered bisemifermions.

\item bring together the ``~$J$~'' bisemifermions in such a way that the off-diagonal bisemisheaves
\[ ( T_{ST_{L_{R_i}}} \widetilde M^c_{\o{ST}_{\o v_{\sigma _i,m_{\sigma _i}}}}
\times T_{ST_{L_j}} \widetilde M^c_{{ST}_{v_{\sigma _j,m_{\sigma _j}}}}) \;, \quad \forall\ i\neq j\;, \]
of interaction of the ``~$ST$~''  structures, but also of the ``~$MG$~'' and ``~$M$~'' structures, have a continuous character.
\Ee
\vskip 11pt

\section{Proposition}

{\em {\bfseries An isomorphism $I^c_{QT\to GR}$ exists between the string field equations of the internal dynamics of a set of compactified bisemiparticles and the equations of general relativity\/}:
\begin{multline*}
I^c_{QT\to GR} : \quad 
[(T_{ST_{L_{R_i}}}\times T_{ST_{L_j}})(\widetilde M^c_{\o{ST}_{\o v_{\sigma _i,m_{\sigma _i}}}}
\times \widetilde M^c_{{ST}_{\o v_{\sigma _j,m_{\sigma _j}}}})]\\
\begin{aligned}
+ & [(DT_{MG_{L_{R_i}}}\times DT_{MG_{L_j}})(\widetilde M^c_{\o{MG}_{\o v_{\sigma _i,m_{\sigma _i}}}}\times 
\widetilde M^c_{{MG}_{v_{\sigma _j,m_{\sigma _j}}}})]\\
& \quad =-[(DT_{M_{L_{R_i}}}\times DT_{M_{L_j}})(\widetilde M^c_{\o{M}_{\o v_{\sigma _i,m_{\sigma _i}}}}\times 
\widetilde M^c_{{M}_{v_{\sigma _j,m_{\sigma _j}}}})]
\end{aligned}\\
\To \lambda g_{\mu \nu }+G_{\mu \nu }=8\pi GT_{\mu \nu }\end{multline*}
at the conditions that \cite{Pie1}:
\Bean
\item  the sections of the right and left semisheaves ``~$ST$~'', ``~$MG$~'' and ``~$M$~'', which are in fact ``~$ST$~'', ``~$MG$~'' and ``~$M$~'' strings, be viewed as families of geodesics which must be interpreted as the flow lines of a fluid.

\item the directional gradients used in the left hand side of $I^C_{QT\to GR}$ be replaced by covariant derivatives on the right hand side.
\Ee
}
\vskip 11pt

\bpr
\Be
\item At first, the conditions a) will be precised.

The sections of the semisheaves  ``~$MG$~'' and ``~$M$~'' are assumed to correspond respectively to families of geodesics $\Ps_{MG}(\lambda _{MG},n_{MG})$ and $\Ps_{M}(\lambda _{M},n_{M})$~, where:
\Bi
\item $\lambda _{MG}$ and $\lambda _M$ are affine parameters telling where we are on a given geodesics,
\item $n_{MG}$ and $n_M$ are selector parameters allowing to distinguish one geodesics from the next \cite{M-T-W},
\Ei
if these sections have been desingularized in order that the tangent vectors
\[ \vec u_{MG}=\F{\partial \Ps_{MG}}{\partial \lambda _{MG}} \quad \and \quad
\vec u_{M}=\F{\partial \Ps_{M}}{\partial \lambda _{M}}\]
be parallel on the corresponding geodesics.

Note that the vectors
\[ \vec n_{MG}\equiv \F{\partial \Ps}{\partial n _{MG}} \quad \and \quad
\vec n_{M}\equiv \F{\partial \Ps}{\partial n _{M}}\]
measure the separation between points with the same values $\lambda _{MG}$ and $\lambda_M$ on neighbouring geodesics.

\item Then, the terms ``~$ST$~'', ``~$MG$~'' and ``~$M$~'' will be shown to be in one-to-one correspondence on the left and on the right of $I^c_{QT\to GR}$~.

\Be
\item The term $(T_{ST_{L_{R_i}}}\times T_{ST_{L_j}})(\widetilde M^c_{\o{ST}_{\o v_{\sigma _i,m_{\sigma _i}}}}
\times \widetilde M^c_{{ST}_{v_{\sigma _j,m_{\sigma _j}}}})$~, describing the internal vacuum structures ``~$ST$~'' at the level of elementary particles and being of expanding space-time nature (which can be seen by the differential nature of the bioperator
$(T_{ST_{L_{R_i}}}\times T_{ST_{L_j}})$ acting on $H^+_{ST}$ according to section 3.10), must correspond to the term $\lambda g_{\mu \nu }$ of $GR$ if it is taken into account that:
\Bi
\item this term $\lambda g_{\mu \nu }$ is not very well shaped, all the information having been smashed in the cosmological constant $\lambda $~.

\item the internal vacuum energy density $\rho ^{(ST)}_{\rm vac}$~, to which $\lambda /8\pi G$ corresponds, must be of expanding space-time nature: for this reason, the space-time differential operator $(T_{ST_{L_{R_i}}}\times T_{ST_{L_j}})$ was chosen and not a directional gradient bioperator of the type $(DT_{ST_{L_{R_i}}}\times DT_{ST_{L_j}})$ as introduced in proposition 3.11.
\Ei

\item The term $(DT_{MG_{L_{R_i}}}\times DT_{MG_{L_j}})(\widetilde M^c_{\o{MG}_{\o v_{\sigma _i,m_{\sigma _i}}}}\times 
\widetilde M^c_{{MG}_{v_{\sigma _j,m_{\sigma _j}}}})$~, describing the ``middle ground'' vacuum structures ``~$MG$~'' of elementary particles and being of contracting space-time nature, is in one-to-one correspondence with the term $G_{\mu \nu }$ of the general relativity equations if:
\Bi
\item the conditions (a) and (b) of this proposition are taken into account.
\item it is noted that each right section
$\widetilde M^c_{\o{MG}_{\o v_{\sigma _i,m_{\sigma _i}}}}$ has been projected onto each left section $ 
\widetilde M^c_{{MG}_{\o v_{\sigma _j,m_{\sigma _j}}}})$ in such a way that they are confounded: they can then be rewritten according to:
\[ (\widetilde M^c_{\o{MG}_{\o v_{\sigma _i,m_{\sigma _i}}}}\times 
\widetilde M^c_{{MG}_{v_{\sigma _j,m_{\sigma _j}}}})
\To
\widetilde M^c_{{MG}_{v_{\sigma _{i-j},m_{\sigma _{i-j}}}}}\;.\]
\Ei
So, to the set $\{\widetilde M^c_{{MG}_{v_{\sigma _{i-j},m_{\sigma _{i-j}}}}}\}$ of bisections of ``~$MG$~'' will correspond a family $\Ps_{MG_{L_R-L}}(\lambda _{MG},n_{MG})$ of products of left geodesics, localized in the upper half space, by projected symmetric right geodesics localised in the lower half space, in such a way that to the term 
$(DT_{MG_{L_{R_i}}}\times DT_{MG_{L_j}})
(\widetilde M^c_{{MG}_{v_{\sigma _{i-j},m_{\sigma _{i-j}}}}})$ will correspond the term $\vec\nabla _{\vec u_{MG}}\centerdot \vec \nabla _{\vec u_{MG}}\centerdot \vec n_{MG}$~.

$\vec \nabla _{\vec u_{MG}}\  \vec n_{MG}$ is the covariant derivative of the vector field $ \vec n_{MG}$ along a product, right by left, $\Ps_{MG_{L_R-L}}(\lambda _{MG})$ of symmetric geodesics with tangent vector
\[ \vec u_{MG}=\F{\partial \Ps_{MG}}{\partial \lambda _{MG}}\;, \]
and
$\vec\nabla _{\vec u_{MG}}\  \vec \nabla _{\vec u_{MG}}\  \vec n_{MG}$ is the corresponding relative acceleration.

As the relative acceleration of geodesics allows to define the Riemann curvature tensor \cite{M-T-W} by:
\[ \vec\nabla _{\vec u_{MG}}\  \vec \nabla _{\vec u_{MG}}\  \vec n_{MG} + \text{Riemann}(\ldots
\vec u_{MG}, \vec n_{MG},\vec u_{MG})=0\]
which leads to the components of the tensor of Riemann $R^\lambda _{\mu \lambda \nu }$ in a coordinate basis.

Due to the antisymmetric property of $R^\lambda _{\mu \lambda \nu }$ \cite{Wei1}, there are only two tensors which can be generated by contraction from $R^\lambda _{\mu \lambda \nu }$~: it is the Ricci tensor $R_{\mu \nu }\equiv R^\lambda _{\mu \lambda \nu }$ and the curvature scalar $R=R^\mu _\mu $ \cite{Dar}.

So, the only tensor which can be formed from $R_{\mu \nu }$ and $R$ is the tensor $G_{\mu \nu }=R_{\mu \nu }-\half g_{\mu \nu }R$ which is thus in one-to-one correspondence with the ``~$MG$~'' term on the left hand side of $I^c_{QT\to GR}$~.

\item Finally, the term
$(DT_{M_{L_{R_i}}}\times DT_{M_{L_j}})(\widetilde M^c_{\o{M}_{\o v_{\sigma _i,m_{\sigma _i}}}}\times 
\widetilde M^c_{{M}_{v_{\sigma _j,m_{\sigma _j}}}})$~, describing the ``mass'' structures of elementary particles and being also of contracting space-time nature, is in one-to-one correspondence with the tensor $T_{\mu \nu }$ of the equations of general relativity.

Indeed, if it is taken into account that the ``mass'' term ``~$M$~'' and the
``middle ground'' term ``~$MG$~'' on the left of $I^c_{QT\to GR}$ have the same structure, it is immediate to associate with
$(DT_{M_{L_{R_i}}}\times DT_{M_{L_j}})(\widetilde M^c_{\o{M}_{v_{\sigma _{i-j},m_{\sigma _{i-j}}}}})$ the ``mass'' curvature tensor of Riemann $R^\lambda _{ \mu \lambda \nu }$ from which the ``mass'' tensor
\[ G_{\mu \nu }^{(M)}=R^{(M)}_{\mu \nu }-\txt \half g^{(M)}_{\mu \nu }R^{(M)}\]
can be formed.

Considering the Einstein equations
\[ G_{\mu \nu }=8\pi GT_{\mu \nu }\;, \]
it is evident that:
\Be
\item[(1)] the tensor $G^{(M)}_{\mu \nu }$ is equal to the ``mass'' energy-momentum tensor $8\pi GT_{\mu \nu }$~.
\item[(2)] to the mass term on the left hand side of $I^c_{QT\to GR}$ corresponds the tensor $T_{\mu \nu }$~.\epr
\Ee
\Ee\Ee
\vskip 11pt

\section{Proposition}

{\em In the context of the new interpretation of the equations of general relativity describing the dynamics of the generation of the mass shells of the elementary particles from their vacuum structures, we have that:
\Be
\item the cosmological constant will likely have a high value and will be noted $\lambda _{ST}$~.
\item the internal vacuum structure ``~$ST$~'' itself (without the generated ``~$MG$~'' and
``~$M$~'' structure) at the elementary particle level is probably responsible for the dark energy.
\Ee}
\vskip 11pt

\bpr
\Be
\item The cosmological constant $\lambda _{ST}$~, referring now to the internal vacuum structures of elementary particles, is now directly related to the Planck scale; consequently, $\lambda _{ST}$ will have a high value.

\item The dark energy, being of expanding space-time nature, will probably correspond to sets of elementary particles endowed only with their internal vacuum structures ``~$ST$~''.\epr
\Ee
\vskip 11pt

\section{The algebraic quantum theory is a quantum gravity theory}

The algebraic quantum theory \cite{Pie5}, recalled at the beginning of this chapter for a set of interacting bisemiparticles, is a quantum gravity theory.

Indeed, the gravity is no more introduced ontologically as resulting from the curvature of space-time but from the diagonal interactions between right and left semisheaves
``~$ST$~'', ``~$MG$~'' and ``~$M$~'' belonging to different bisemiparticles as it was developed in \cite{Pie5}.  We refer thus to the preprint ``Algebraic quantum theory'' \cite{Pie5} for a description of quantum gravity as resulting from the diagonal interactions between pairs of semiobjects.